\documentclass[]{acmart}

\usepackage[group-separator={,}]{siunitx}
\usepackage[normalem]{ulem}
\usepackage{tabularx}

\newcommand{\mlg}[1]{}
\newcommand{\msb}[1]{}
\newcommand{\lp}[1]{}
\newcommand{\wh}[1]{}
\newcommand{\joon}[1]{}
\newcommand{\zxm}[1]{}
\newcommand{\jt}[1]{}

\newcommand{\revdelete}[1]{}

\newcommand{\citem}[1]{\mbox{\cite{#1}}}

\AtBeginDocument{%
  }

\setcopyright{none}
\acmJournal{PACMHCI}
\acmYear{2023} \acmVolume{7} \acmNumber{CSCW2}
\acmArticle{337} \acmMonth{10} \acmPrice{15.00}
\acmDOI{10.1145/3610186}

\received{January 2023}
\received[revised]{April 2023}
\received[accepted]{May 2023}

\begin{document}

\title{Cura: Curation at Social Media Scale}

\author{Wanrong He}
\affiliation{%
  \institution{Tsinghua University}
  \city{Beijing}
  \country{China}
  }
\email{hewanrong8@gmail.com}

\author{Mitchell L. Gordon}
\affiliation{%
  \institution{Stanford University}
  \city{Stanford}
  \country{United States}
  }
\email{mgord@cs.stanford.edu}

\author{Lindsay Popowski}
\affiliation{%
  \institution{Stanford University}
  \city{Stanford}
  \country{United States}
  }
\email{popowski@stanford.edu}

\author{Michael S. Bernstein}
\affiliation{%
  \institution{Stanford University}
  \city{Stanford}
  \country{United States}
  }
\email{msb@cs.stanford.edu}

\renewcommand{\shortauthors}{Wanrong He et al.}

\begin{abstract}
How can online communities execute a focused vision for their space? Curation offers one approach, where community leaders manually select content to share with the community. Curation enables leaders to shape a space that matches their taste, norms, and values, but the practice is often intractable at social media scale: curators cannot realistically sift through hundreds or thousands of submissions daily.
In this paper, we contribute algorithmic and interface foundations enabling curation at scale, and manifest these foundations in a system called \textit{Cura}.
Our approach draws on the observation that, while curators' attention is limited, other community members' upvotes are plentiful and informative of curators' likely opinions.
We thus contribute a transformer-based curation model that predicts whether each curator will upvote a post based on previous community upvotes. Cura applies this curation model to create a feed of content that it predicts the curator would want in the community.
Evaluations demonstrate that the curation model accurately estimates opinions of diverse curators, that changing curators for a community results in clearly recognizable shifts in the community's content, and that, consequently, curation can reduce anti-social behavior by half without extra moderation effort.
By sampling different types of curators, Cura lowers the threshold to genres of curated social media ranging from editorial groups to stakeholder roundtables to democracies.\footnote{The code is available at \url{https://github.com/StanfordHCI/Curation-Modeling}.}

\end{abstract}



\begin{teaserfigure}
  \centering
  \includegraphics[width=1\textwidth]{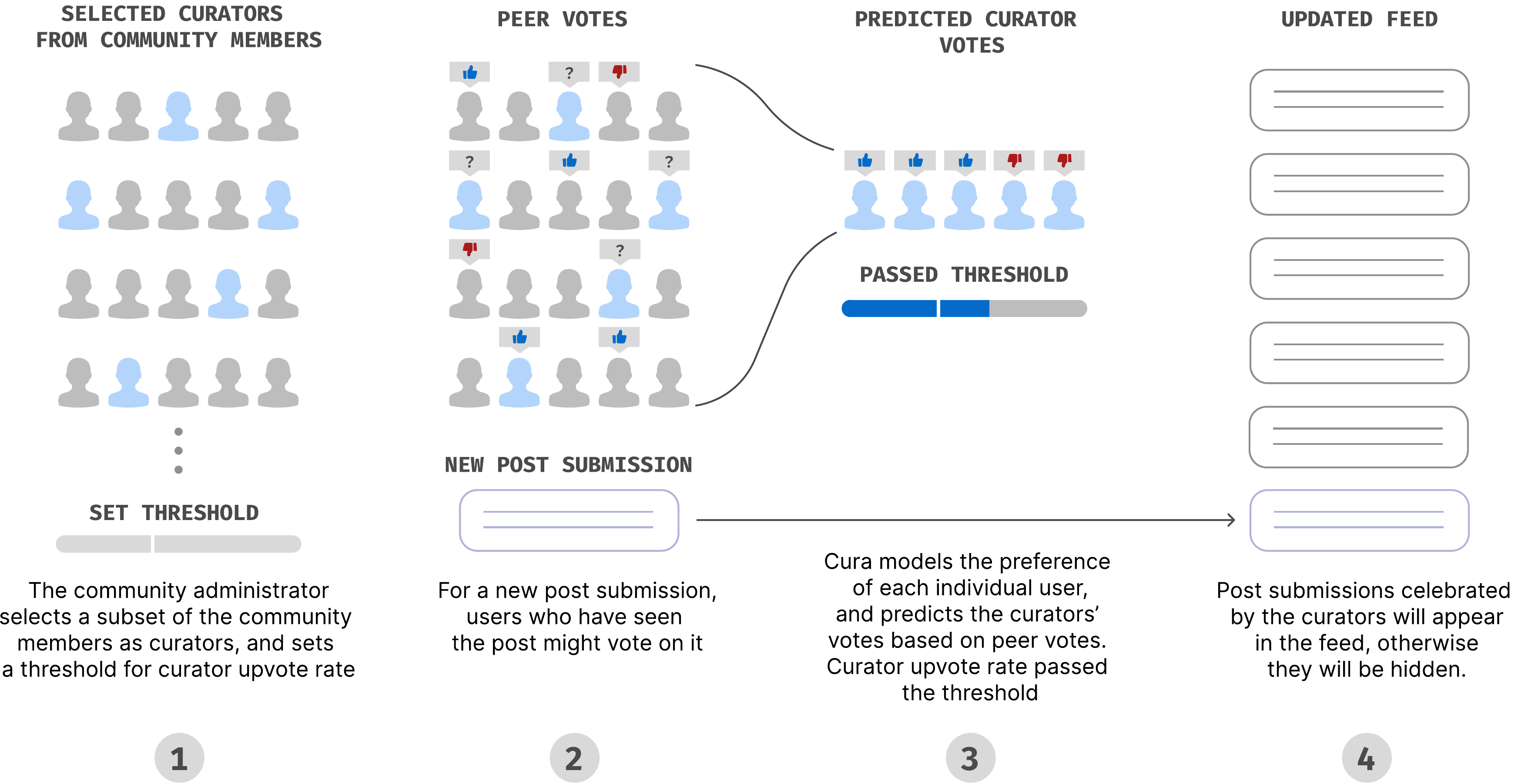}
  \caption{Cura empowers communities to create feeds that reflect the taste and opinions of designated \textit{curators, from specific individuals to a democratic polity.} (1) The creator selects members as curators whose upvote behavior should determine the content that should be included in the community. (2) Given other community members who have voted on a post, (3) Cura estimates the curators' preferences for the post submission. If the proportion of curators who upvoted or are predicted to upvote passes a threshold, (4) the post appears in the group feed; otherwise, it remains backstage.} %
  \Description{This figure contains four columns. The first column displays 20 members of a community, fifteen people in grey, and five in blue, indicating they are the selected curators. A thermometer in grey with a bar on the two-fifth position is under the 20 people, where the bar indicates the threshold for the curator upvote rate set by the community. In the second column, a thumb-up/thumb-down emoji is on the top of a subset of people, representing they have upvoted/downvoted on the new post submission. Under the people is a text box in purple, which represents the post submission. The third column displays five people in blue, the curators, and three of them have thumb-up emojis while two have thumb-down emojis on top, representing their own vote or the predicted vote by our Cura model. Under the people is the same thermometer, but three-fifths of it is covered in blue and has passed the bar, indicating the curator upvote rate is above the threshold. The fourth column displays a column of text boxes in black with the newly added purple text box at the bottom, indicating that the post submission is curated and appears in the feed.
  }
  \label{fig:teaser}
\end{teaserfigure}

\maketitle

\section{Introduction}
Curation enables us to execute focused visions for social spaces. Curated social spaces arise both online and offline: editorial boards for The Atlantic and the New York Times curate which submitted op-eds to publish, administrators at popular social news sites such as Slashdot manually select a small number of submitted tech news stories per day to publish~\citem{lampe2004slash}, museum curators decide on which pieces of art to showcase and arrange into shows, teachers pick examples of student work to share with the class, and online publishers decide which news or comments to highlight~\citem{diakopoulos2015picking,wangcurationasmoderation}. As seen in these examples and others, a curation metaphor empowers community leaders to select which content is shared with the community. By giving curators control over what community members see, curation empowers curators to strongly influence the group's descriptive norms~\citem{rajadesingan2020quick,cheng2017anyone,cialdini1991focus,kraut2010dealing}.

Curation, however, has remained intractable in many large-scale online communities. How might a community leader or moderator~\citem{seering2022metaphors} execute a focused vision in a community that may have thousands of members or more, as on Reddit, Facebook Groups, large Slack workspaces or e-mail lists, or forums? Manually reviewing and approving every submitted post can require incredible effort: curating the NYT Picks comments in the New York Times online requires an entire dedicated, paid team~\citem{diakopoulos2015picking,wangcurationasmoderation}. Already, moderators' bandwidth on these platforms is limited to reviewing only a very small proportion of content~\citem{park2022measuring}, and moderators are often already overworked~\citem{gillespie2018custodians}. Lacking tools to enable curation in large-scale communities, many platforms today adopt the metaphor of broadcast in which all content and replies are posted into a large public square instantaneously, then rely on the community itself to vote on or otherwise review the submissions. This broadcast metaphor has been successful because it lowers the effort threshold~\citem{myers2000past,ross1991person} to sharing and consumption, allowing communities to gain and keep attention at scale~\citem{chen2021}. Unfortunately, broadcast also facilitates off-topic and aggressive posts that attract attention~\citem{hsu2021social} and erode pro-social norms~\citem{chang2016engineering,cheng2017anyone,kiene2016surviving}, unraveling the goals that a curated community might seek.

To expand the availability of curation to large-scale online communities, in this paper we introduce a system called \textit{Cura}.
Cura's goal is to facilitate online communities whose content reflects the curators' opinion on what to share, and can do so at scale without curator review on every submission.
Our approach is motivated by the observation that \textit{community members'} opinions are plentiful and can provide substantial information about \textit{curators'} opinions. 
By observing which community members upvote a post, we can learn to predict whether the curators will upvote the post for curation into the community.
This approach allows communities to run at scale on user contributions, while maintaining a north star of their curators' opinions.
We operationalize this approach by contributing a deep learning transformer architecture, which we call a \textit{curation model}, that takes as input the target curator(s), the community, the post, and any upvotes on the post from community members, then estimates whether the curator will upvote the post.\footnote{We use upvotes as the predicted curator behavior in this paper, but this action could be replaced with any other behavior: e.g., moderation decisions, likes, favorites, retweets, comments, or emoji reactions.} Learning patterns between community votes and curator votes, the curation model updates a post's status as it observes new upvotes from community members.
We additionally contribute interactive models in Cura: a management interface for curated communities, where the community's administrators or moderators can select curators whose taste the system should model; a frontstage for posts that clear the model's curation threshold; and a backstage for the others where they are less visible but still browsable to the community for further feedback.

We evaluate Cura on existing Reddit data to test whether it accurately estimates curators' opinions, empowering the creation of different spaces by selecting different curators. We sample curators from a variety of identity and political affiliations via a large public dataset of upvotes on Reddit. We find that a curation model trained on this Reddit dataset estimates curators' votes with accuracy $81.96\%$, whereas the traditional majority vote (Reddit `score') performs at $65.9\%$.
We also observe that the curation model also estimates present-but-quieter users---lurkers as curators---accurately, enabling curation practices that more reflect the broader community and not just the most active members.
We then demonstrate that applying curation to large and popular subreddits can execute dramatic shifts, for example refocusing a large technology community around human-centered concerns, or removing anti-LGBTQ+ content from a large teenagers community.
In a field experiment, we confirm that these shifts are clear and recognizable to people: participants can accurately identify which versions of a community were curated in our system by specific subgroups.
Finally, we measure that curation roughly halves the number of anti-social violations in a large community---without requiring any additional moderation effort.

Curation facilitates the creation of social media community types at scale that would be otherwise difficult.
The contents of today's social media are influenced most heavily by the upvote behaviors of the most active users. But, by selecting different curators, communities can embed stronger opinions about the kind of behaviors that are welcomed. Curators might represent a small group of editors (a la a small hosted workshop). Or, by selecting \textit{all} community members as curators, the community might behave as a democracy: one that intentionally includes active lurkers who visit often but vote less often, by predicting their opinions as well. Or, a community might establish a curation group that is a stakeholder roundtable~\citem{gordon_jury_2022}, ensuring that minority voices are heard, unlike a straight majoritarian vote. By making explicit and transparent who is guiding decisions in a community, curation may facilitate deliberate discussion in each community about what sort of space they want to inhabit.

In summary, this paper contributes:
(1)~the Cura system to empower curation on large volumes of content without continuous manual review by drawing on community members' behavior; the system consists of both 
(2)~a transformer-based curation model that predicts curators' opinions---whether they will upvote or downvote---based on community feedback, and (3)~curation interface patterns such as frontstage and backstage spaces that allow community feedback to power the curation model's input; the model and interface are realized in (4)~a functional implementation of these on the Reddit platform. Through this work, we aim to %
concretely demonstrate the feasibility of curation at social media scale.

\section{Related Work}

\subsection{Curation, moderation, and norms in social computing}

Curation, in a social computing context, denotes the process of a curator selecting a subset of the content that fits their preferences or the demand of a larger community, where the content can be news articles, user submitted posts, or comments~\citem{seering2022metaphors,wangcurationasmoderation}. %
An early study of e-mail list moderators lists reports that they would sometimes engage in curation decisions about what content to allow, claiming that ``quality must be maintained or the audience will desert the group''~\citem{berge_perceptions_2000}. Such tensions persist today~\citem{zhang2015mailing}. %
Many individual users proactively curate for themselves, e.g., by following certain users or groups~\citem{lee2019consumptive}, blocking certain sources~\citem{Naderer2020TheSA}, and searching~\citem{Khatter2020AnalysisOC}. Users productively curate for other users as well, e.g., by reframing and sharing content with designated groups of people~\citem{lee2019consumptive, Lisa-Block-2021, Park2019MediatingRO}. Some users also curate to meet their self-expressive and self-representational needs~\citem{Buss2021TransgenderIM, villi_social_nodate}. This self-curation is a critical part of an effective social media diet, though it has less of a direct influence on the wider community.

Curation for social media communities arises in some modern web contexts: \citeauthor{diakopoulos2015picking} notes that a small subset of comments at the New York Times are flagged as `NYT Picks'~\citem{diakopoulos2015picking}, and \citeauthor{Bruns2010} identifies that Kuro5hin and Plastic allow users to vote on submitted stories before they are published~\citem{Bruns2010}. Even at a limited scale, these curation approaches require extensive human effort: in the case of the New York Times, it requires a dedicated paid team. Our objective is to develop a system that can support this sort of curation at scale without overwhelming or overly bottlenecking on human curators and the time that community members need to wait for approvals.  %

On the other hand, the term `algorithmic curation' in previous literature mostly refers to content personalization for individual users based on machine learning algorithms~\citem{Swart-Experiencing-2021,Gausen2022UsingAM}. Such personalization algorithms create different experiences for each community member, so they are not community-level curation, and have also been implicated in increasing polarization and creating echo chambers~\citem{Gausen2022UsingAM}. In contrast to this prior work, our goal is to facilitate group-level curation, so our approach models a single decision criterion (the curator's taste) on behalf of the entire community.  

The practice of upvoting and downvoting content by users and subsequently ranking it according to their votes on social media platforms such as Reddit can be viewed as a form of crowdsourced curation at a group level. However, it limits the pool of curators to the active members of the community. Therefore, our objective is to broaden the scope of curators and open up more possiblities for curation, for example to encompass lurkers, or to focus on a specific set of community members.

Curation is also related to the practice of moderation. Moderation is typically focused on the task of removing objectionable or norm-violating content: as Gillespie puts it, ``both to protect one user from another, or one group from its antagonists, and to remove the offensive, vile, or illegal''~\cite[p.~5]{gillespie2018custodians}. Much of the labor of moderators is focused on this sort of removal~\citem{roberts2014behind,li2022all}.
Curation, as a form of moderation, can be separate from removal: as \citeauthor{wangcurationasmoderation} put it, ``highlighting high-quality content [is] a moderation strategy''~\citem{wangcurationasmoderation,gillespie2018custodians}. Inductive work has identified `curator' as a metaphor that moderators may use to describe themselves~\citem{seering2022metaphors}. Under the grammar of moderation~\citem{grimmelmann2015}, including principal techniques of moderation (nouns) and important distinctions in how moderation is carried out (adverbs), we can describe this existing curation practice as a manual, ex ante, centrally executed organization of content. 
In this paper, we draw on this practice to imagine a version of curation that can scale to larger communities, instead being described by the grammar as automatic, ex ante, and distributed in nature.
Such curation has yet to be supported by a socio-technical system.
Similar to those moderation tools that are designed for content deletion and removal at scale, we envision curation to be a collaborative effort of human curators with the support of algorithmic tools~\citem{gillespie2018custodians,seering2019moderator}.

Curation is not the only method of shaping norms in an online community. Community members respond to and replicate the behavior they see around them~\citem{rajadesingan2020quick,chang2016engineering}, learning and following what are known as descriptive norms~\citem{kiesler2012regulating}. In broadcast, this makes violations visible to all members, and the violations tend to accrue engagement~\citem{hsu2021social}, creating a negative spiral in which more and more norm-violating behavior may arise over time~\citem{cheng2017anyone}. Making norms salient can increase adherence amongst community members~\citem{cialdini1991focus}, for example by including permanent notes reminding members of the rules when they are authoring new content~\citem{matias2019preventing}. Identity signals can also influence norms, for example anonymity vs. pseudonymity~\citem{kiesler2012regulating} as well signaling ingroup identity on the platform~\citem{seering2018applications}. By giving curators control over the ways in which designed signals are portrayed to the community, curation seeks to engage similar mechanisms.

\subsection{Algorithms supporting social media}

We draw on advances in machine learning and social computing to enable curation. Most machine learning classifiers only model one voice, the ground truth it has learned from the dataset, often reflecting the majority voice of the dataset labelers~\citem{gillespie2018custodians, schaekermann_ambiguity-aware_2020, rogers_changing_2021,gordon2021disagreement}. Such a classifier can not satisfy the need for different communities with distinct tastes and values, and can result in the ``tyranny of the majority'' by ignoring minority groups' viewpoints~\citem{gordon_jury_2022}. 
Researchers have been calling on the development of algorithms that can balance the needs and values of multiple stakeholders and achieve collective goals (e.g.,~\citem{abebe_roles_2020, sasha_design_2020, dobreski_toward_2018, moreau_paradigm_2019, smith_keeping_2020, yu_keeping_2020, zhu_value-sensitive_2018}).
We draw inspiration from jury learning~\citem{gordon_jury_2022}, which learns to predict how each individual user would label unseen examples to enable selection of which voice the community wants to listen to. We extend this concept to learn from other community members' votes in order to make a more accurate prediction.

AI models on social media such as Twitter and TikTok draw on insights from recommender systems to personalize content based on individual users' preferences. Our approach similarly make personalized predictions and draws on insights from recommender systems: we develop a user embedding as part of the curation model, allowing the model to make vote predictions for any user in its training data. Our model, however, is expected to update its prediction with each new vote from a member of the community.  %
Researchers on recommender systems have articulated the need for value-sensitive design~\citem{stray_building_2022, chen_practitioners_2022}, particularly to integrate human values and explicitly optimize the recommender systems for higher measurements on those values~\citem{stray_building_2022}.
Our approach to this issue is to ask each community to make an explicit value statement through its decision of which members serve as curators, and which posts those curators upvote.

Collaborative filtering is a group of popular algorithms for building recommender systems which leverage the known preferences of a group of users to predict the unknown preferences of other users~\citem{su_survey_2009,Resnick1997RecommenderS}. 
Memory-based collaborative filtering, which calculates the similarity between items or users based on common users or items~\citem{koren_advances_2022}, is widely-deployed in commercial recommender systems~\citem{amazon_CF, Hofmann2004Latent}. However, it suffers from sparsity issues when there are few common items, and is hard to scale for large datasets~\citem{su_survey_2009}. %

Particularly salient for our context is the cold start problem: collaborative filtering cannot make predictions on new items. %
Content-based recommender systems help overcome this problem by leveraging features of the users and items for making predictions, where the features include categorical features like genre, content, and URL domain~\citem{su_survey_2009, ContentBasedPazzani, koren_advances_2022}. 
In our work, we draw on content features from natural language processing~\citem{shalom_natural_2022}. Such techniques include Long Short-Term Memory~\citem{hochreiter_long_1997}, or transformers~\citem{Transformer} such as BERT~\citem{devlin-etal-2019-bert}, a pretrained model designed to understand natural language text. %
These architectures can be used to model categorical attributes of users and items, as well as model users' historical behaviors---a sequence of items the user has interacted with~\citem{sun_bert4rec_2019, zou_improving_2022}. %
We exploit this capability of BERT by directly inputting the features of the user, the community, and the post as tokens toward prediction.

Under curation, only posts that are celebrated by the curators are published, preventing the spread of norm-violating posts as well as the resulting harm. In contrast to many algorithic moderation tools which directly measure the content toxicity, an algorithmic curation tool would attempt to forecast the reception of content by a particular curator or group of curators. Researchers have already developed models for forecasting conversational events---what a piece of content might bring up, whether it will be perceived as helpful~\citem{althoff-etal-2016-large} or lead to derailment~\citem{chang-danescu-niculescu-mizil-2019-trouble,schluger2022proactive}, steps toward the proactive, discerning application of curation that we envision.

\section{Cura: Design and Interaction}

In this section, we describe our design goals of the curation metaphor and the system, \textit{Cura}, that we built to realize it.

\subsection{Motivating Scenario}\label{sec:scenario} %

Several years ago, Alice created an online community about technology news because of her interests in security and regulation. It started as a group with people who shared her interests, but over time expanded to cover all tech news. Eventually, Alice and the original cohort of members left: the community no longer featured content that they cared about as the tides had turned towards crypto. Alice decides to restart a community around technology security and regulation news, but does not want to experience the same problematic shift again, so she opts to implement curation for this new community. She makes herself and some of the more active participants from the old community into the curators for this new one. This time around, when people join the community, they see news stories around recently discovered security vulnerabilities at the top of the feed. When someone tries to flood the community with thinly veiled cryptocurrency scams, the content remains backstage and not visible to most members, while a story around the SEC considering cryptocurrency regulation does move frontstage, since it fits Alice's vision for the group. %

\subsection{Design goals}

Broadcast in social media prioritizes low friction to sharing content, roughly akin to giving everybody in the audience a microphone. Curation is more akin to a hosted show, where there is greater friction to getting content shared with the entire group, so the host can better shape the conversation. Each has its place in our social ecosystem: broadcast social media for more rapid-fire and free-flowing spaces, and curation for slightly slower spaces that pursue a particular set of norms.

Curation already occurs in some communities online, but it is limited by the sheer amount of effort required to pre-review every submission in large-scale communities. This project explores a design space in which curators loosen the reins slightly to allow an algorithm, with oversight, to aid them in curation. They do so by allowing the algorithm to learn from the curators' voting behavior and how it can be predicted by a combination of topic and votes from others in the community---which other community members' votes are predictive of the curator's votes? When the system gains confidence in the curators' likely behavior, it allows the post to move forward. To integrate this model into a system, we introduce interface approaches for curation, including a backstage for content that is not (yet) above the threshold for curation, as well as interfaces for articulating the curators for each community.

We detail here our design goals for our curation system:

\paragraph{(1)~Operate on large-scale communities without overloading curators with work or grinding the community to a halt waiting for approvals.} Moderators are overworked~\cite{roberts2014behind}, and large communities cannot come close to reviewing each piece of content for norm violations. %
If members find that their content is left indefinitely in a review queue, they may stop posting. So, we must thread the needle by identifying content while requiring manageable oversight from curators and only small delays. We achieve this by leveraging active community members and learning the correlations between the curators' opinions and those community members. %

\paragraph{(2)~Identify posts that curators support for their community.} Toward empowering communities to better execute on their desired norms, curation models the curators' opinions on content in the community, and only promotes content that those curators would support. The design insight that motivates our system is that curator attention is scarce, while community attention is more plentiful. So, we draw on signals from community members while maintaining the curators' point of view as a North Star.

\subsection{System interaction} \label{sec:interaction}

\begin{figure*}[tb]
  \centering
  \includegraphics[width=0.65\textwidth]{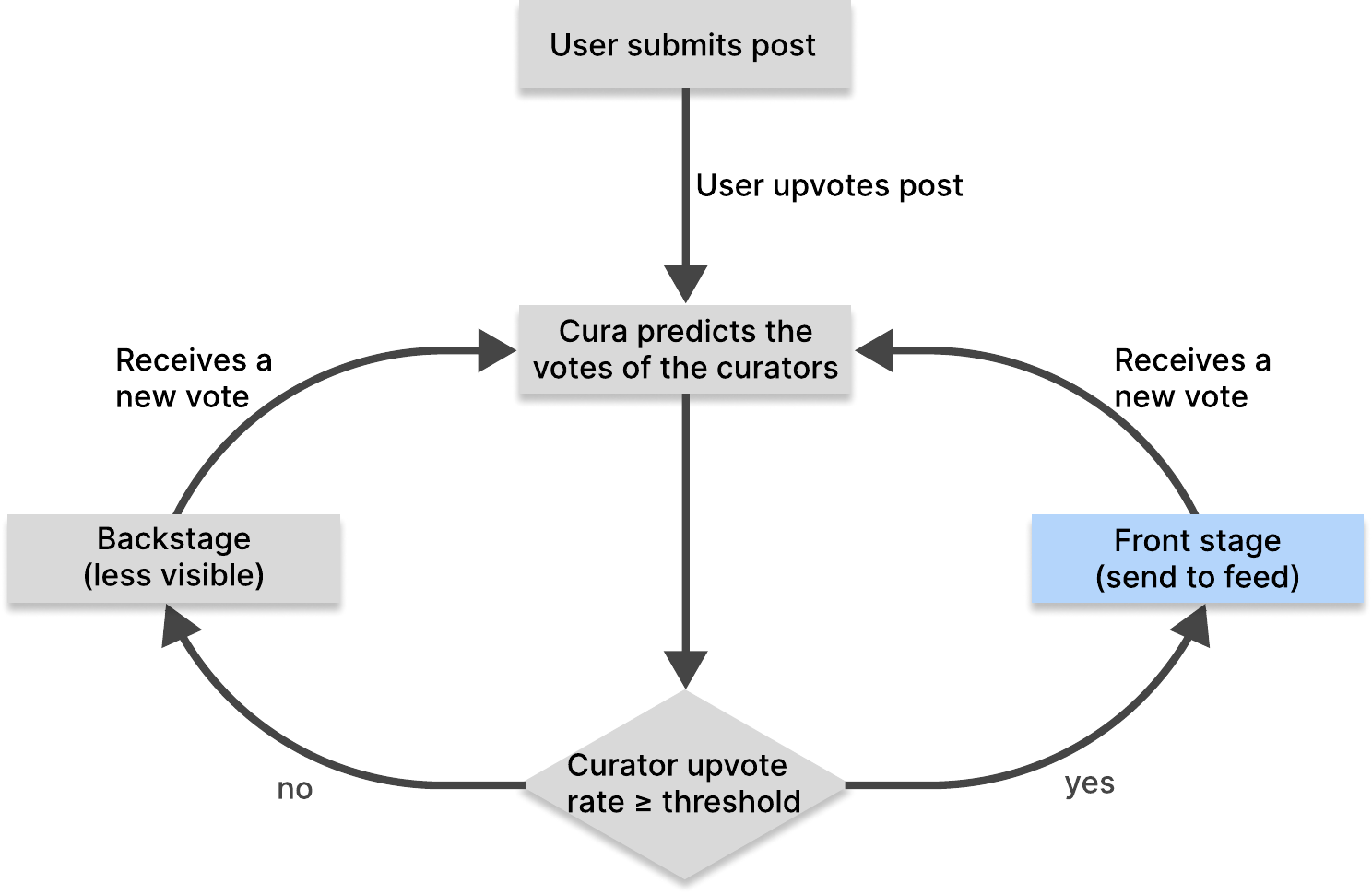}
  \caption{A post's flow through Cura: When a new post arrives, Cura estimates the proportion of curators who would upvote it. If the proportion is above the community's curation threshold, the post goes live in the frontstage feed that is presented to members by default. If the proportion is below the community's curation threshold, it remains in a less visible backstage area which can only be viewed by trusted members or if members actively browse it. When a new vote comes in from a community member, Cura updates its estimates and reroutes the post to the frontstage or backstage appropriately.} %
  \Description{This figure is a flow chart that contains five states: the initial state is "User submits post". An arrow says "User upvotes post" starts from this state and points to the next state that "Cura predicts the votes of the curators". An arrow starts from this state and points to the state "Curator upvote rate greater than or equal to threshold". An arrow that represents the "yes" condition points to the state of the "Front stage", and another arrow that represents the "no" condition points to the "Backstage". Both "Front stage" and "Backstage" are pointed to the state "Cura predicts the votes of the curators" by arrows that say "Receives a new vote".}
  \label{fig:algo_flow}
\end{figure*}

We instantiate these ideas in \textit{Cura}, a curation system for social media. Cura is currently implemented as an alternative feed visualization for Reddit, and is envisioned eventually as a standalone social media site. In the system, a community administrator selects a set of community members (Reddit accounts) as curators for their communities.\footnote{Because our system is not directly integrated with Reddit, Cura does not directly enforce a requirement that it can only be used by administrators. We envision that, if this were a standalone social media platform, it would be available to community administrators.} The interaction for this process is similar to selecting moderators. The administrator then selects a curation threshold that determines when a post gets selected to share with the group. For example, a curation threshold of 50\% implies that 50\% of the subreddit's curators must either manually vote to upvote the post, or be estimated to upvote the post, in order to share it. As the administrator selects curators, Cura supports them by visualizing information about the curator's history and activity levels (\autoref{fig:prac_interface}). Members browsing the subreddit continue to upvote or downvote posts as before according to their own preferences, and posters continue to submit posts as before. The key difference is that these upvotes and downvotes are now used to predict curator behavior: once the curators are defined for a community, Cura draws on signals from the post content, curator voting history, and upvotes and downvotes from community members to estimate whether each curator would upvote the content. If the curators' estimated approval clears a threshold, the post is shared in the community's public feed. Further votes can update the prediction, even hiding the post again if it falls below threshold. The system learns which community members are highly correlated with the curators and which are not, so it can ignore trolls, but update quickly with informative votes. \autoref{fig:algo_flow} illustrates the flow of the system. Below, we detail the different interfaces through which users interact with Cura.

\subsubsection*{Selecting curators}
\begin{figure*}[tb]
  \centering
  \includegraphics[width=1.0\textwidth]{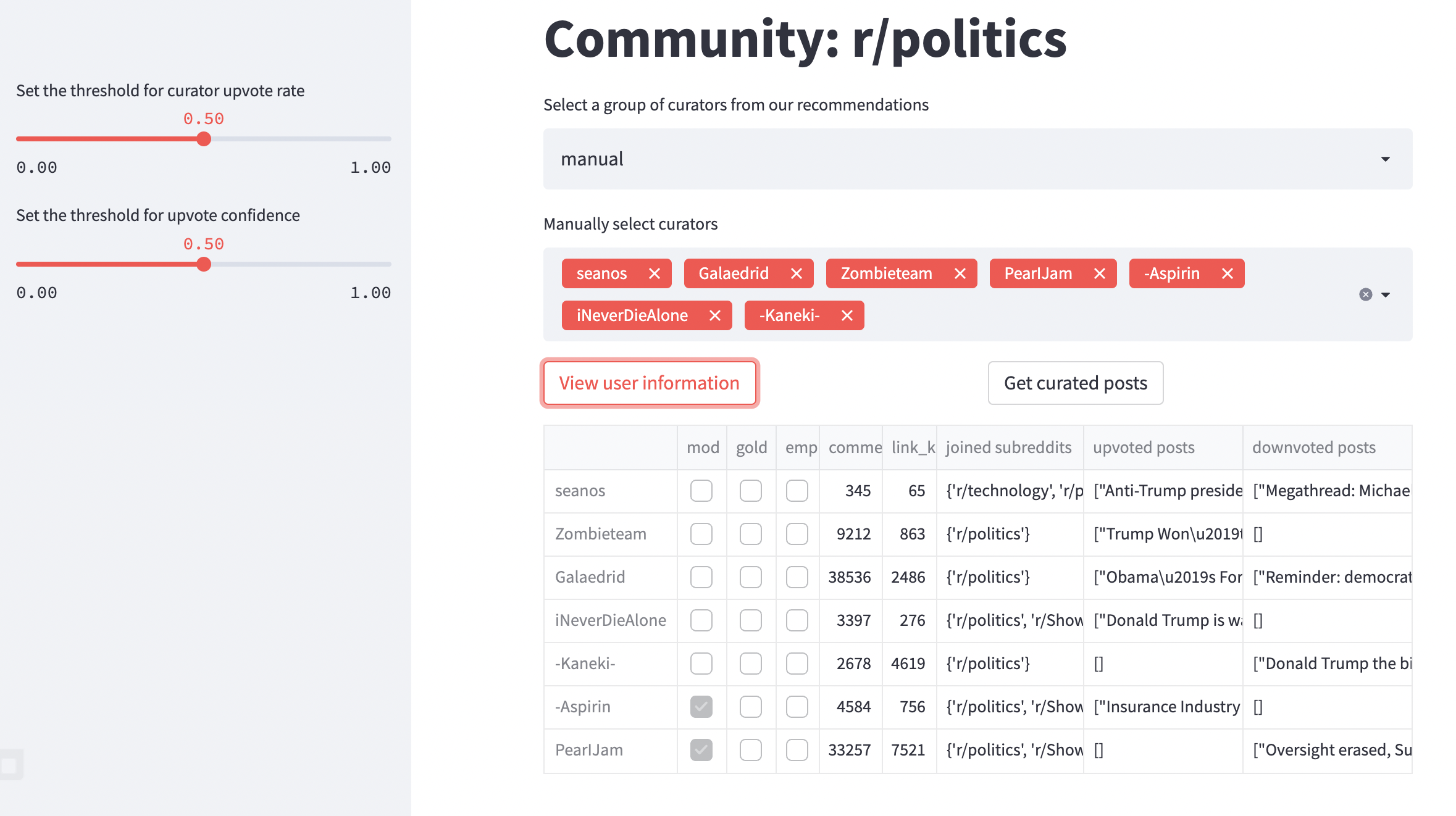}
  \caption{Administrators define the curators on behalf of the community, set the threshold for how many curators should be estimated to upvote a post in order for it to go live to the entire community and the threshold for how much confidence should the estimation have to predict a curator upvote. } 
  \Description{The figure shows the interface for the Cura system. The interface has the title "Community: r/politics". Under the title, there is a drop-down menu where the practitioner can select a group of curators from the system recommendation on behalf of the community. In this figure, the practitioner selects the "manual" option. Underneath is an input box where the practitioner manually selects the users seanos, Galaedrid, Zombieteam, PearlJam, -Aspirin, iNeverDieAlone, -Kaneki- as the curators. The practitioner clicks on the button "View user information," and a table is shown below, which illustrates whether those selected users are mods, Reddit gold, Reddit employees, their link karma, their comment karma, the subreddits they have joined, the posts they have upvoted or downvoted on.}
  \label{fig:prac_interface}
\end{figure*}

Curators are the members of each community whose upvote behavior is emulated for decision-making. In \autoref{fig:prac_interface}, the administrator or moderator uses Cura to define curators for their community. Cura enables them to view information about community members, including the communities they have joined and posts they have upvoted or downvoted, and select a subset of users as curators. In addition, this administrator selects (1)~a curation threshold determining the percent of curators who must support a post (through a combination of actual upvotes and the model's predictions) for the post to go live to the community, and (2)~a confidence threshold above which Cura believes that a curator will support a post. The existence of two orthogonal thresholds exists primarily for expert use: for most cases, the first one suffices. The second threshold exists in case the administrator needs to make the model more or less conservative in its predictions---if it would be preferred, for example, for the model to only grant a curator's upvote when it is extremely certain. As an example of how combining these thresholds affords additional configurability for experts when needed, an administrator might state that the model must be 90\% confident to predict that an individual curator will upvote in favor of curating a post into the community, and require at least half of curators to have voted in favor in order for the post to move ahead.  %

How might a community select curators? Curators might be selected based on their interests and opinions, identity, or experience and past behaviors in the community.  %
In Cura currently, we leave control of the curator selection to the community via the administrator, so that Cura can be guided by the community's values. We envision that curators might eventually be selected through election~\cite{zhang2020policykit}, named by trusted individuals such as administrators, or created de facto by the community purpose (e.g., consider the celebrity of a fan community). In the Discussion section, we reflect on how this selection process centralizes the power in the hands of the curators, and the governance processes that might aid it or prevent communities from coopting this approach for harm. In our design, curators are envisioned to be community members who are trusted by their peers, and selected based on factors such as expertise, opinion diversity, and identity diversity (\autoref{sec:governance}).

By selecting curators, communities create different genres of community curation. A small group of selective curators might create a space with a highly specific purpose and strident sense of taste, and a more democratic community might select every member in the community as curators: not just the most active contributors, but also those who are often present but rarely vote (e.g., lurkers). %
The only technical requirement is that curators must be active enough for Cura to learn patterns in their behavior, e.g., at least $5$ votes in the community. In the absence of this minimum engagement, there exists a potential risk of low-quality user modeling.

\subsubsection*{Generating the community's feed}

\begin{figure*}[tb]
  \centering
  \includegraphics[width=1.0\textwidth]{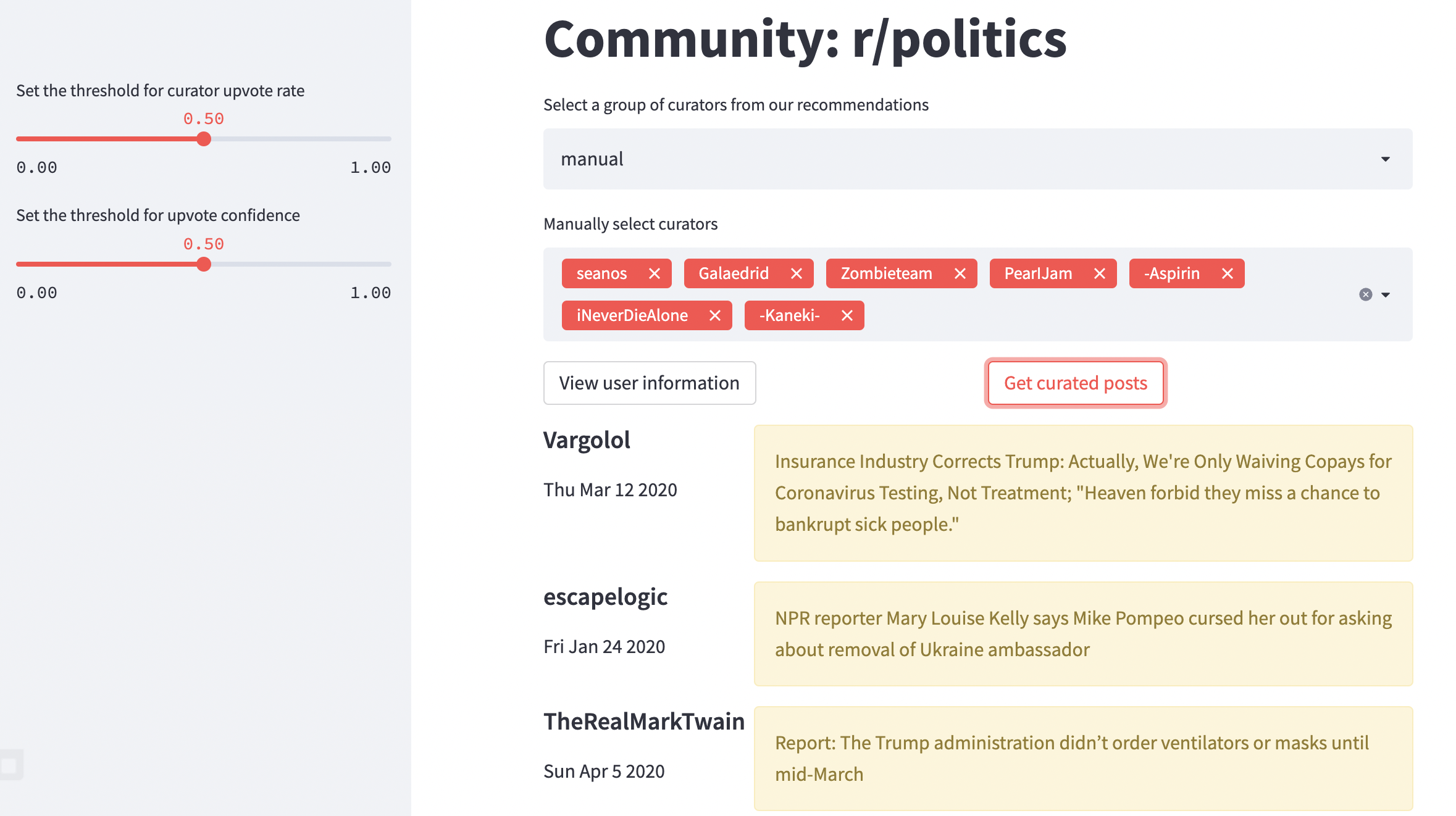}
  \caption{Cura enables exploration of community feeds based on the selected curators and the set curation threshold. Here, Cura presents a version of Reddit's r/politics community that is filtered based on a set of seven curators.} 
  \Description{The figure shows the interface for the Cura system. The interface has the title "Community: r/politics". Under the title, there is a drop-down menu where the practitioner can select a group of curators from the system recommendation. In this figure, the practitioner selects the "manual" option. Underneath is an input box where the practitioner manually selects the users seanos, Galaedrid, Zombieteam, PearlJam, -Aspirin, iNeverDieAlone, -Kaneki- as the curators. The practitioner clicks on the button "Get curated posts" and the curated posts, or the feed, are shown below, containing the posts from the users Vargolol, escapelogic, and TheRealMarkTwain. There are two sliders on the left sidebar. The practitioner set the threshold for curator upvote rate as 0.5 using the first slider and the threshold for upvote confidence as 0.5 using the second slider.}
  \label{fig:cura_feed}
\end{figure*}

We refer to the posts that clear the curation threshold as the community's \textit{feed} (\autoref{fig:cura_feed}). We recommend setting the curation threshold as $50\%$, so that a post needs to receive a majority of upvotes from the curators to be featured in the feed. The administrator can experiment with different thresholds and evaluate their impact on the community's feed through the interface (\autoref{fig:cura_feed}). Subsequently, the curation threshold can be dynamically adjusted to align with the administrator's preferences.

\autoref{tab:teenager_curator_content} demonstrates several different feeds curated from the r/teenagers subreddit community, where the curators of the first four feeds are the intersection of the active users in r/teenagers and the active users in r/Feminism, r/MensRights, r/LesbianActually and r/gay, respectively. The fifth group of curators is 50 randomly chosen users in r/teenagers. The feeds can be strikingly different from each other, with the r/teenagers feed curated by users in r/Feminism and r/LesbianActually promoting more woman-centric viewpoints. Choosing to empower a particular set of curators effectively also means choosing to diminish the influence of conflicting community members: the r/MensRights feed is clearly different from that curated by r/Feminism, for example, but both of these are viewpoints represented today within the broadcast r/teenagers community. Curation gives a community the chance to choose whose voices they want in control when there are conflicting values.%

\begin{table}[tb]
  \footnotesize
  \def\arraystretch{1.5}
  \begin{tabular}{p{2.43cm}p{2.43cm}p{2.43cm}p{2.43cm}p{2.43cm}} 
  \toprule
  \textbf{r/Feminism members curate r/Teenagers} & 
  \textbf{r/MensRights members curate r/Teenagers} & 
  \textbf{r/LesbianActually members curate r/Teenagers} & 
  \textbf{r/gay members curate r/Teenagers} & 
  \textbf{Random members of r/Teenagers curate r/Teenagers} \\ \midrule
  Some girl bragged about having a relationship for 5 months. Jokes on her! I've been with loneliness for nearly 17 years &
  My crush was complaining about not having a boyfriend and then I said “I could be your boyfriend” She laughed &
    Your cat isn't a "Thicc Boi" it's fucking overweight &
    No one fucking loves me my parents are abusive and i constantly get fucking bullied because i am “the quiet kid” &
    Ok so who’s joining? \\ 
  PSA: This subreddit is not a dating app, you horny teens &
    I think there’s a squatter in my attic &
    iCarly was the first e girl. prove me wrong &
    Just so proud &
    Being the smart kid is a stress \\
  Happy Alentines Ay for those of us not getting the V or D &
    For all you teens out there that are in "that" mood, here you go &
    So does that mean I'm in a relationship? &
    Those bastards lied to me &
    Why do they always scream. \\
  My boyfriend broke up with me, I got a 72 on an exam, and my dad had a heart attack. Happy Valentine’s Day! &
    My country is fucked.(UK) Government says 60\% of the population needs to be infected with the corona virus to develop hers immunity &
    i am very fancy &
    R.I.P kobe. He was my legend, he will always live on in my heart. Still cant belive he actuelly died, so sad :( &
    He is now known as Steve from Accountin \\ \bottomrule
  \end{tabular}
  \caption{Feeds from the r/teenagers subreddit community, created with Cura by changing the set of curators for r/teenagers. Selected posts were those only preferred by one group of curators or that ranked substantially higher in one curated feed than others by curator upvote rate.} %
  \label{tab:teenager_curator_content}
\end{table}

Cura's curation threshold determines the percent of curators who must be estimated to upvote the content or actually upvote the content. %
Table~\ref{tab:teenager_threshold} demonstrates the r/Teenagers feed curated by all community members who have voted $\geq 5$ times under increasingly strict thresholds. As the threshold increases, the content is more pro-social, but the risk increases of a false negative in the model prediction. Cura enables the practitioner to iteratively explore different compositions of curators and different thresholds to see how they will influence the resulting feed (\autoref{fig:prac_interface}). 

\begin{table}[tb]
  \footnotesize
  \def\arraystretch{1.5}
  \begin{tabular}{p{3.12cm}p{3.12cm}p{3.12cm}p{3.12cm}}
  \toprule
  \textbf{r/Teenagers curation threshold of 80\%} & \textbf{r/Teenagers curation threshold of 60\%} & \textbf{r/Teenagers curation threshold of 40\%} & \textbf{r/Teenagers curation threshold of 20\%} \\ \midrule
  For the two people who wanted it! An update to my artwork &
    My cat hurt his foot a few days ago. To lift his spirits, let's make him into a meme format. &
    Almost lost my cool there &
    Any hot grills wanna have a chat \\
  My dog Teddy got put down today, so I drew him as a last goodbye &
    Like I did the best mom,give me my ps4 back &
    Boys who don't like them selves and think their ugly. If you would like to answer, why? &
    can someone please crush my skull with a hammer? im having delusions about everything \\
  You dare use my own spells against me, Potter? &
    saw this kid at the cafeteria browsing r/teenagers, make him see himself &
    Too many virgins on this sub reddit lol &
    Just chatting with the GFs sister \\
  Same mirror, but one year apart! Anorexia recovery has been hard but seeing these pictures makes me proud of how far I've come &
    A doctor just flirted with me today. She told me that I am too sweet. Well her exact words were "severely diabetic" but I know what she meant. &
    Unpopular Opinion: heterophobic. Nobody is born straight. Strait people shouldn't adopt because straight people raise straight kids. &
    Driving tip for all you newbies \\ \bottomrule
  \end{tabular}
  \caption{Curated feed for the r/teenagers subreddit community with every community member being a curator. Each column shows post submissions with the curator upvote rate above a certain threshold but under a higher threshold, i.e., post submissions in the columns to the left are also above the threshold for this column.} %
  \label{tab:teenager_threshold}
  \end{table}

\begin{figure*}[tb]
  \centering
  \includegraphics[width=0.95\textwidth]{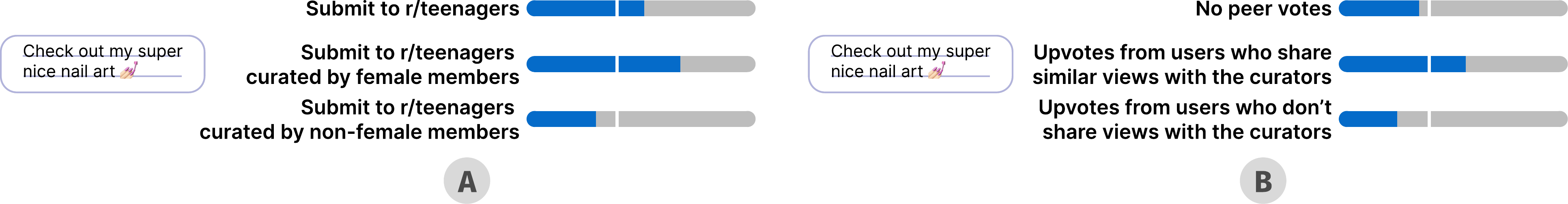}
  \caption{Conceptual visualizations of how the same post might fare via curation across different communities and different voting groups. (A)~Changing the set of curators affects whether posts on teen girls' and womens' issues clear the curation threshold. (B) If a post is upvoted by members who have not historically shared opinions with the curators, it is not curated as above threshold into the community. } %
  \Description{Both figure A and figure B have two columns. The first column in figures A and B is a text box representing a post submission that says, "Check out my super nice nail art". The second column in figures A and B contains three thermometers with bars at the same position representing the same threshold for the curator upvote rate, and blue stripes representing the temperature, which is also the actual curator upvote rate. 
  In figure A, the first thermometer has the note "Submit to r/teenagers" and the temperature is slightly above the bar. The second thermometer has the note "Submit to r/teenagers curated by female members" and the temperature is high above the bar. The third thermometer has the note "Submit to r/teenagers curated by non-female members" and the temperature is slightly below the bar.
  In figure B, the first thermometer has the note "No peer votes" and the temperature is slightly below the bar. The second thermometer has the note of "Upvotes from users who share similar views with the curators" and the temperature is above the bar. The third thermometer has the note of "Upvotes from users who don’t share views with the curators" and the temperature below the bar.}
  \label{fig:changing_threshold}
\end{figure*}

An end-user visualization displays how close each post is to clearing the curation threshold for that space (\autoref{fig:changing_threshold}A). Given the same number of upvotes, the outcome may differ depending on which community members upvoted the post (\autoref{fig:changing_threshold}B). 

\subsubsection*{Backstage}
We introduce a backstage in Cura in support of content that is not yet above the curation threshold. If integrated into existing platforms such as Reddit that keep all posts visible, an implementation of Cura as described so far could use the curation estimate from 0\% to 100\% as a ranking score. This would enable a version of curation that integrates into a feed ranking algorithm, or be translated into a curator upvote visualization (\autoref{fig:changing_threshold}).

However, as a standalone platform, our instantiation of curation in Cura enables a split of content between the \textit{backstage} and \textit{frontstage}. Posts live in the backstage before the curation algorithm is confident in their quality, where community members may manually visit and vote to update the prediction; by default, however, members see the curated frontstage feed. When the algorithm updates its prediction for the curators to be above the chosen threshold, posts move to the frontstage where the whole community can see them.

Backstaging enables community input---so the curators do not need to manually review every post---while keeping norm-violating content and anti-social behaviors in a low status and less visible area, mitigating potential damage to the community from norm degradation and harassment. If less visibility is desired, the community administrator could only give access to the backstage to trusted community members. Just as Reddit identifies promising content based on a few users actively browsing the ``new'' feed, a standalone version of curation social media such as Cura will rely on a few community members exploring the backstage and upvoting content to bring it close to or above the curation threshold.
Submitting a post to a community operating with Cura is akin to submitting a paper to a conference, where some posts may never make it to the frontstage. However, users have the option of submitting posts that align more closely with curators' preferences, increasing the chances of appearing on the frontstage.

\section{Technical Approach} %

Our goal is to enable curation at social media scale. To achieve this, we do not require that curators manually review every post. We thus seek to model curators' votes based on other members' votes---in other words, predicting ``If members X and Y upvoted the content, and member Z downvoted it, what is the curator likely to say?'' We also require an infrastructure that can update its prediction after each new peer vote, so that content can dynamically move between the frontstage and backstage. 

To achieve these goals, we present a transformer-based model, which we call a \textit{curation model}. Our curation model builds on the recent success of transformers~\cite{Transformer} to create a transformer to classify whether the target user will upvote a post. The input to the model is a concatenation of the target user (curator) and stringified features of the post, including the post text content, post author, community where the post is published, voting time, whether the post is flagged as NSFW,\footnote{``Not Safe For Work'', a flag used to label content that contains explicit or offensive material that may be inappropriate for a workplace or public setting.} and the URL domain linked by the post.
Following prior work~\cite{Fan2022BuildingMQ, Beltagy2020LongformerTL}, we use special tokens before each feature to indicate the type of the feature being encoded. We represent each user as a unique token so that they have a unique embedding in the model. For example, to predict at ``Nov 6, 2023'', whether user $a$ will upvote ``Glaciers in Europe are experiencing the most severe melting on record'' submitted by the user $b$ in a community named ``News'', the converted model input string will be ``\texttt{[USERNAME] [$a$] [AUTHOR] [$b$] [COMMUNITY] News [CREATED\_TIME] Wed Nov 6, 2023 [NSFW] false [SUBMISSION\_URL\_DOMAIN] www.washingtonpost.com [SUBMISSION\_TEXT] Glaciers in Europe are experiencing the most severe melting on record}''. %

Our implementation of this curation model uses a BERT encoder~\cite{devlin-etal-2019-bert} with a linear projector for binary classification.
Encoding the input with the BERT encoder, the linear projector takes the encoding of the target user token as input and outputs a scalar $p$, which represents the probability of the target user upvoting on the post. $p$ indicates the confidence of the model prediction. 
We obtain the final decision---whether the target user will upvote---by thresholding the probability $p$, e.g., at a default value of $0.5$. %
Users can adjust the threshold to control the composition of the front-stage content (\autoref{fig:prac_interface}).

The curation model draws on others' voting behavior to estimate the curator's vote. To achieve this, we train on history votes from peers in the communities\footnote{Votes on content from all communities contribute to the model training and prediction. While different communities may have varying voting norms, the curation model is capable of learning how users' voting behavior differs from one community to another based on the community name included in the model input.}, including the initial upvote from the author themselves (\autoref{sec:interaction}). The model receives training supervision through a series of training examples, for example, that Users X and Y upvoted post P, User Z downvoted post P, and the curator upvoted post P. From this, it learns to leverage prior upvotes or downvotes in its training data via the overlapping signals on post P to predict the vote for a given user. 
Then, to make a prediction informed by current upvotes, when the Cura system receives a new vote on a post, we finetune the full curation model (including the BERT encoder and the linear projector) one step further on the new vote, then re-predict for each curator.
This finetuning adjusts the model's prediction of the curator's opinion, creating a system that is responsive to community feedback. Additionally, it allows the model to place greater emphasis on recent votes over older ones, which facilitates the adaptation of the model to changes in users' opinions. Even if the model initially makes incorrect predictions on users' changing opinions, it can learn from users' votes and capture their latest preferences over time. Finetuning is rapid enough for interactive purposes with sub-second speeds; if faster speeds are needed, parts of the model could be frozen during finetuning.

\subsection{Dataset}
For training and evaluation, we draw on a dataset of Reddit upvotes and downvotes from accounts that voluntarily opted in to make their vote data public. On Reddit, users could navigate to their profile and opt in to make their voting history public, including the posts they voted on and whether they upvoted or downvoted the post.\footnote{This preference is visible on the ``old Reddit'' profile page, and still functions for users who click through to it. The preference is called ``make my votes public''. The newer Reddit design de-prioritizes this preference. The resulting dataset is at: \url{https://www.reddit.com/r/help/comments/8x11lp/how_to_make_upvoted_public/} We utilize a dataset that collected these opt-in users' history voting data.\footnote{``Huge Collection of Reddit Votes'', \url{https://www.kaggle.com/datasets/josephleake/huge-collection-of-reddit-votes}}} 
We use the following post features provided by the Reddit dataset to compose the model input: unique Reddit username, voting time, subreddit, post author Reddit username, and whether the post is flagged as NSFW. We additionally use PRAW,\footnote{\url{http://praw.readthedocs.io}} the Python Reddit API wrapper, to retrieve the URL domain linked by the post (if present\footnote{Only posts of the ``link'' type contain a retrievable URL domain. Other post types will result in an empty string.}), as well as the text content of the post. %

Due to computational resource limits, in this paper we only use a subset of this full vote Reddit dataset. We sample every vote on posts in a set of popular subreddits on the topics of politics (r/politics, r/Conservative, r/Liberal, r/Republican, r/democrats, r/VoteBlue), jokes (r/Jokes, r/Showerthoughts), science (r/science, r/ScienceFacts, r/technology, 
r/shittyaskscience), and gender, sexual orientation, and identity (r/Feminism, r/MensRights, r/LesbianActually, r/gay, r/trans) and lifestyle (r/teenagers).\footnote{These subreddits were chosen due to their representation of distinct identities, interests, and views. The selection allows us to conduct experiments, as outlined in \autoref{sec:diff_curators}, whereby we designate users from certain subreddits as curators. We aim to confirm that the curation outcomes differ significantly depending on the curators selected.}
 
This sample results in \num{1966122} votes, which comprise 4.4\% of the dataset, containing \num{518798} different posts. We randomly split the dataset (votes) into training and test sets at a 80\%--20\% ratio. %
This voting data, collected from the real-world Reddit behavior, is very unbalanced: 74\% of the votes are upvotes and only 26\% are downvotes, and 10\% of the most active users account for 86\% of the votes while most of the other users vote rarely and are more typically lurkers. Our model design and training must account for these imbalances. 

\subsection{Training}\label{sec:training}

We use an implementation of the BERT transformer model from Huggingface~\cite{huggingface}. We initialize the weights of the BERT encoder with the pre-trained BERT-mini~\cite{bert-mini}, then expand the model vocabulary and token embedding by adding special indicator tokens for each feature and adding a unique token for each user. The embedding of the added tokens and the weights of the linear projector are initialized randomly. We then train our model to predict user votes by fine-tuning on our Reddit dataset.

To address the problem of unbalanced voting data, we apply a weighted binary cross entropy loss. For a user-post-vote triplet where user $a \in U$ voted $x\in V=$ \{upvote, downvote\} on post $s\in P$, where $U$ is the set of users and $P$ is the set of posts, we assign the weight to be $$W(a, s, x) = \frac{1}{\left|\{(u, p, v)| u = a, v = x\}\right|} * \frac{\left|\{(u, p, v)| p = s\}\right|}{\left|\{(u, p, v)| p = s, v = x\}\right|} * w(x)$$
where upvotes receive $w(x)$ of 1, and downvotes 1.5.
The weight is inversely proportional to the number of up- and downvotes the user has made, and inversely proportional to the proportion of users that have the same opinion $x$ on post $s$ as the target user $a$. This weighting ensures that the model prediction is not biased towards the most active users or toward the majority opinion.
We add an additional weight for downvotes to balance the model's strong tendency of predicting upvotes.

We trained our model on one Tesla V100 GPU for 10 epochs with a batch size of 64 and a maximum text length of 512 tokens. We use Adam~\cite{Kingma2015AdamAM} optimizer with a learning rate of $3e-5$. When the Cura system receives new votes, we adopt a learning rate of $3.6e-5$ for finetuning. The hyperparameters are selected after a small grid search.

\section{Evaluation}

This paper proposes that the Cura system can reshape the contents of a community, and that curators' opinions can be estimated effectively based on community members' opinions. In this evaluation, we test these proposals by mapping them onto two main questions: (1)~Does our model estimate curators' votes accurately? (2)~What impact does curator selection have on the algorithm's selected posts for a community?

\subsection{Estimating Curators' Votes}
Whether Cura is able to accurately predict curators' opinions on posts is a core question and is the cornerstone of other claims. To answer this question, we test whether our model can accurately predict votes from arbitrary users---essentially, any curator we might select---in the test set of our Reddit vote dataset. We use upvotes as proxies for content that these users actively want to see in the community, so our goal is to predict upvotes accurately.

\subsubsection{Performance across users}

\begin{figure}[tb]
  \centering
  
  \includegraphics[width=0.45\textwidth]{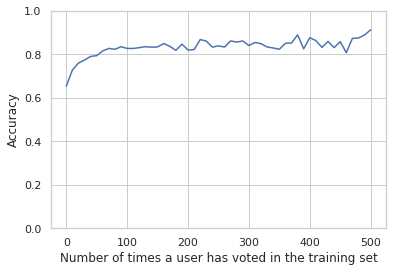}
  \includegraphics[width=0.45\textwidth]{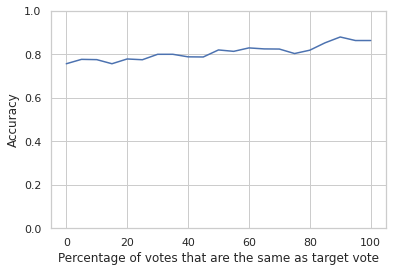}
  
  \caption{(A) For active users who vote more than fifty times, the model achieves $80\%$ accuracy. The model requires few votes to begin modeling users accurately, improving from $62\%$ for users who have voted only once to $73\%$ for users who have voted ten times. (B) Accuracy for target votes that are the same with different proportions of votes: the accuracy is always above $70\%$ no matter whether the vote is the majority or minority opinion.}
  \Description{Figure A is a line chart where the x-axis is the number of times a user has voted in the training set, and the y-axis is the accuracy. The accuracy gradually increases as the number of votes increases when there are less than 50 votes, and becomes relatively stable at around 85%
  }
  \label{fig:user_num_acc}
\end{figure}

To calibrate performance measures, we sampled a subset of upvotes from the original test set to form a balanced test set, where the number of upvotes is equal to the number of downvotes for each subreddit, so that random guessing is $50\%$. 
The balanced test set consists of \num{127528} posts, with \num{103818} upvotes and \num{103818} downvotes in total. Each post corresponds to 1.63 votes on average, ranging from one vote to 52 votes, with a standard deviation of 1.88 and a median of one vote.
Perfect accuracy of $100\%$ is not feasible, given that people often hold self-inconsistent opinions and would not react the same way if they saw the same post a second time~\cite{callison2009fast,gordon2021disagreement}. %

Our model achieves an accuracy of $81.96\%$, with a ROC AUC score of $0.8903$. For comparison, we trained an ablated baseline model that predicts the majority vote for each post. When testing on the same balanced test set, this baseline achieves $65.96\%$ accuracy and a ROC AUC of $0.7241$, indicating that our model better predicts curator votes than a majority vote. %

\begin{table}[tb]
    \begin{tabular}{l|ll}
    Actual/Predicted & Upvote & Downvote \\ \hline
    Upvote           & 85039 (0.8191)  & 18779 (0.1809)    \\
    Downvote         & 18673 (0.1799)  & 85145 (0.8201)   
    \end{tabular}
    \caption{Confusion matrix of the vote prediction result on the balanced test set. The model achieves a high accuracy for both upvotes and downvotes.}
    \label{tab:confusion}
\end{table}

Curators are likely active users, and the model performs at over $80\%$ accuracy for active users. However, some communities might want to implement democracy as a curation strategy, which would also require estimating lurkers' opinions so that it can proxy their votes as well. How well does the model estimate lurkers' votes? Accuracy begins at $62\%$ for users with only one vote in the training data, and raises to $73\%$ after ten votes (\autoref{fig:user_num_acc}(A)). 

Does the model actually learn the preferences of individual users, or is it copying majority vote? The model estimates both majority ($\geq 50\%$ same votes) and minority ($\leq 50\%$ same votes) opinions with roughly equal accuracy (\autoref{fig:user_num_acc}(B)). This indicates that the model is learning individual opinions rather than ``piling on'' majority votes.

\subsubsection{Performance across topics}
Does the model perform worse in certain communities? \autoref{tab:subreddit_acc} reports accuracy across each of the subreddits in our balanced test set. Performance generally is stable, with a mean prediction accuracy of $77.86\%$ ($\sigma = 8.43\%$). The only low-performance outlier, r/ScienceFacts, has only \num{137} datapoints in the full dataset (108 datapoints in the training set)---too few to model effectively. Based on these results, we suggest having at least \num{2000} datapoints in a community's training set to ensure effective model performance.

\begin{table}[tb]
    \small
  \begin{tabular}{@{}lrrrrrrrrrrrrrrrrrr@{}}
  \toprule

& r/politics & r/Conservative & r/Republican & r/Liberal & r/democrats  \\ \midrule
Number of datapoints & \num{1082006} & \num{63132} & \num{7245} & \num{2076} & \num{6890}  \\
Accuracy  & $84.49\%$ & $91.51\%$ & $88.34\%$ & $79.10\%$ & $85.10\%$  \\ \bottomrule

& r/VoteBlue & r/Showerthoughts & r/Jokes & r/science & r/ScienceFacts  \\ \midrule
Number of datapoints & \num{7732}  & \num{252455} & \num{138451}& \num{111857} & \num{137}  \\
Accuracy & $78.04\%$ & $78.46\%$ & $77.01\%$ & $76.31\%$ & $50.00\%$   \\ \bottomrule

& r/technology & r/shittyaskscience & r/Feminism & r/MensRights  & r/gay   \\ \midrule
Number of datapoints & \num{102203} & \num{7992} & \num{5482} & \num{24682} & \num{4257}  \\
Accuracy & $77.39\%$ & $76.67\%$ & $74.20\%$ & $79.12\%$  & $77.52\%$   \\ \bottomrule

 & r/trans & r/LesbianActually & r/teenagers  \\ \midrule
Number of datapoints  & \num{2597} & \num{2483} & \num{144444}  \\
Accuracy & $69.35\%$  & $81.76\%$ & $77.58\%$     \\ \bottomrule

  \end{tabular}
  \caption{The model performs similarly across a wide variety of subreddits with different number of datapoints in the full dataset.}
  \label{tab:subreddit_acc}
  \end{table}

\subsubsection{Influence of peer votes}
Peer votes should be able to influence the prediction confidence and even correct previously incorrect predictions. Even if the model estimates that a curator is not likely to upvote a piece of content, if several members whose opinions are correlated with the curator's opinion begin upvoting the content, the curator's estimated behavior should shift. To test this effect, we trained a model using the same method from \autoref{sec:training} except that the training set and test set are divided so that there are no overlapping posts between the training set and the test set.
Then, we randomly select one user-post-vote triplet in the test set as the target, randomly shuffle all other votes on the post in the test set and then feed those peer votes to the model one by one. For each new peer vote, we finetune the pre-trained curation model one step further, then evaluate the model prediction on the target vote.

\begin{figure}[tb]
  \centering
  
  \includegraphics[width=0.33\textwidth]{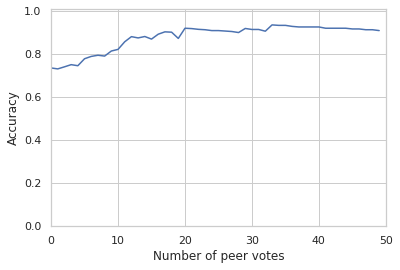}
  \includegraphics[width=0.33\textwidth]{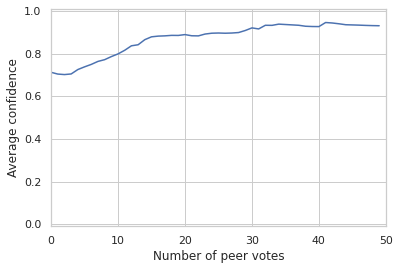}
  \includegraphics[width=0.33\textwidth]{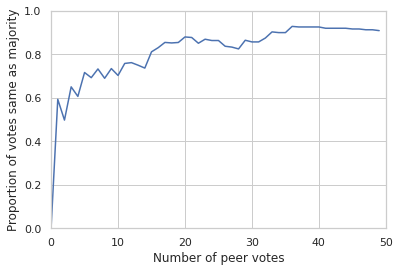}
  
  \caption{(A)~Accuracy increases rapidly as more peer votes are provided, and plateaus around twelve votes. (B)~Confidence also increases as more peer votes are provided. (C)~The model's estimated vote for a curator reacts as more and more peers vote.}
  \Description{Figure A is a line chart where the x-axis is the number of peer votes, and the y-axis is the accuracy. The accuracy gradually increases from 73.03%
  Figure B is a line chart where the x-axis is the number of peer votes, and the y-axis is the average confidence. The average confidence gradually increases from 0.802 to 0.921 for the first 30 peer votes and remains relatively stable.
  Figure C is a line chart where the x-axis is the number of peer votes, and the y-axis is the proportion of votes the same as the majority. The proportion gradually increases from 60%
  }
  \label{fig:peer_vote}
\end{figure}

As the model receives more and more votes from other users, both prediction confidence and accuracy increase (\autoref{fig:peer_vote}): after receiving 10 peer votes, the average confidence of accurate predictions increases from $0.802$ to $0.854$, and the prediction accuracy increases from $73.03\%$ to $82.19\%$. The accuracy further goes up to $92.02\%$ after 20 peer votes. This increment is not a result of copying the majority vote from peers (\autoref{fig:peer_vote}C): %
the model insists on its own prediction when it considers the target user to hold a different view from the majority.

A second test: the model should update its prediction when users with similar preferences as the curator vote. %
We constructed two fake peer vote datasets: we randomly select one user-post-vote triplet in the test set as the target, and then for that user, find the top 50 other users whose voting vector (vector composed by votes on the post in the training set) is most similar to the target user via cosine similarity. We use these voters to construct two datasets where those similar users vote (1)~the same or (2)~the opposite as the target vote. 

\begin{figure}[tb]
  \centering

  \includegraphics[width=0.33\textwidth]{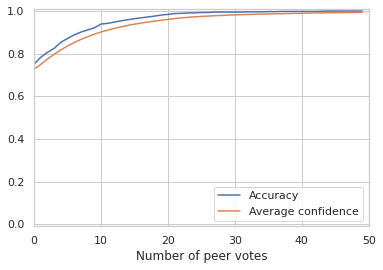}
  \includegraphics[width=0.33\textwidth]{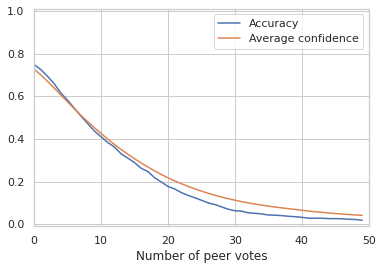}
  
  \caption{(A) When peers with known similar preferences vote, the model adjusts its estimate of the curator's opinion quickly. (B) When peers with known similar preferences vote against the expected outcome, the model loses confidence in its initial vote and eventually flips its outcome.
  }
  \Description{Figure A is a line chart that contains two lines where the x-axis is the number of peer votes, and the y-axis is the accuracy for one line and the average confidence for the other. The two lines almost overlap. The average confidence and accuracy quickly increase from 0.71 and 73%
  Figure B is a line chart containing two lines: the x-axis is the number of peer votes, and the y-axis is the accuracy for one line and the average confidence for the other. The two lines almost overlap. The average confidence and the accuracy gradually drop from 0.71 and 73%
  }
  \label{fig:peer_vote_support_adv}
\end{figure}

Accuracy and the confidence in predictions increase more rapidly as a function of the number of votes when the votes come from similar peers (\autoref{fig:peer_vote_support_adv}(A) vs. \autoref{fig:peer_vote}). The accuracy increasing quickly indicates that the model adapts to peer votes quickly when the peers hold similar tastes. Conversely, if the peers are voting differently, because these peers are especially similar to the curator, the model loses confidence and eventually flips its prediction to the opposite of the ground truth, dropping prediction accuracy to nearly zero (\autoref{fig:peer_vote_support_adv}(B))---as desired. 
However, since the peer votes contradict the model's known preferences of the users, the model is more reluctant to change its prediction than in \autoref{fig:peer_vote_support_adv}(A).%

\subsection{Different Curators Result in Different Feeds} \label{sec:diff_curators}

Given the same inventory of post submissions, do different compositions of curators result in different algorithmically selected feeds? %
To answer this question, we compare curation results between different groups of curators.%

\paragraph{Case study: r/politics.}
We mixed the posts from the subreddits ``r/politics'', ``r/Conservative'', ``r/Liberal'', ``r/Republican'', ``r/democrats'' and ``r/VoteBlue'' by randomly sampling 500 posts from each subreddit to create a super-politics community with a large inventory of politics-related posts that reflect different points of view and norms. The users of this community are the union of the users of the individual subreddits. 

To select curators representing different points of view, we selected users who consistently upvoted posts in each of the constituent subreddits (except the more general r/politics community), specifically those who have voted $\geq 5$ times in that subreddit and have a $\geq 70\%$ upvote rate for that subreddit. We also randomly sampled users from the supercommunity as a control group, thus obtaining 6 groups of curators in total.\footnote{Based on this method, groups may have overlapping curators.} We then applied those six sets of curators to the content of the entire super-community with the curation threshold for curator upvote rate being $50\%$, producing six different feeds.

In terms of content decisions, the preferences of right-leaning curator groups largely differ from those of the left-leaning groups (\autoref{tab:curator_pearson}), while random users are in between but closer to the left-leaning groups, given that r/politics is left-leaning in general. We also present samples of the distinctive posts preferred by each group in Table~\ref{tab:politic_curator_content}: they strongly align with the taste of the curators.

\begin{table}[tb]
\setlength{\tabcolsep}{0.8mm}{
  \begin{tabular}{l|rrrrrr} %
   & r/Conservative & r/Liberal & r/Republican & r/democrats & r/VoteBlue & Random users \\ \hline
r/Conservative & 1. \\
r/Liberal & -0.342 & 1. \\
r/Republican & 0.837 & -0.013 & 1. \\
r/democrats & -0.259 & 0.957 & 0.052 & 1. \\
r/VoteBlue & -0.301 & 0.952 & -0.010 & 0.975 & 1. \\
Random users & 0.004 & 0.878 & 0.308 & 0.889 & 0.865 & 1
  \end{tabular}}
  \caption{Curators with different perspectives result in feeds with very different content. This table reports Pearson correlation coefficients of the curator upvote rates on individual posts, across groups of curators from different subreddits.}
  \label{tab:curator_pearson}
\end{table}

\begin{table}[tb]
  \footnotesize
  \def\arraystretch{1.5}
  \begin{tabular}{p{1.8cm}p{2.3cm}p{1.8cm}p{1.8cm}p{2.1cm}p{2.1cm}}
  \toprule
    \textbf{r/politics curated by r/Conservative users} &
    \textbf{r/politics curated by r/Liberal users} &
    \textbf{r/politics curated by r/Republican users} &
    \textbf{r/politics curated by r/democrats users} &
    \textbf{r/politics curated by r/VoteBlue users} &
    \textbf{r/politics curated by random users} \\ \midrule
  Nashville's Convention Center Discriminates &
    The Navy Has Decided To Restore Capt. Brett Crozier &
    Go Back! &
    Obama says White House response to coronavirus has been `absolute chaotic disaster' &
    Former White House ethics director accuses Trump of `laying groundwork' to interfere with presidential election &
    Emotional Schiff Speech Goes Viral, Delighting the Left and Enraging the Right \\
  ``If China knew of the outbreak way before they told WHO and covered up things, this is a crime against humanity'' - says Italian politician Matteo Salvini &
    Fox Business Blaming Stock Market Drop on Sanders Is a Sign of Things to Come—Ignore the possible pandemic, blame the Bern &
    WHO officials praise US leaders on coronavirus pandemic response: Trump is doing `all he can' &
    Trump: Presidents Have Infinite Power but Cannot and Should Not Ever Be Relied On to Do Anything &
    Health Care Workers Stand Up To Anti-Lockdown Protesters In North Carolina &
    `They Are Saving Our Lives': Demand Grows for Grocery Store Employees, Other Frontline Workers to Receive Hazard Pay Amid Coronavirus Outbreak \\
  Nikki Haley backs investigation of WHO's COVID-19 response: America deserves answers &
    `They Are Saving Our Lives': Demand Grows for Grocery Store Employees, Other Frontline Workers to Receive Hazard Pay Amid Coronavirus Outbreak &
    Cuomo: You know who's doing a good job on NY's COVID-19 outbreak? Trump &
    The Navy Has Decided To Restore Capt. Brett Crozier &
    America doesn’t want another Tea Party - Don’t let Fox News fool you. 81\% of Americans do not share the views of anti-quarantine protesters. &
    Sen. Tammy Duckworth spends Veterans Day in Mexico to be with veterans deported by the Trump administration \\\bottomrule
  \end{tabular}
  \caption{Example posts from the politics community preferred by different groups of curators: either only curated by one group, or ranking much higher in one group than in others. %
  }
  \label{tab:politic_curator_content}
  \end{table}

\paragraph{Case study: r/teenagers.}
We next replicate the same result using communities that are more identity based rather than political. We randomly sample 500 posts from the popular subreddit r/teenagers to create the initial inventory of posts for a teenager community. 
Following the same method as before, we selected different sets of curators as users who upvote content both in r/teenagers as well as in r/Feminism, r/MensRights, r/LesbianActually or r/gay. We also randomly select 50 users from r/teenagers as curators, thus creating four groups of curators. Pearson correlations of curator post upvotes across these groups ranged from $0.614$ (e.g., r/gay and r/LesbianActually) to $0.837$ (r/gay and r/Feminism). \autoref{tab:teenager_curator_content} reports posts uniquely preferred by each group.

\paragraph{Case study: r/technology.} %
In a final case study, we aimed to create a technology community that has a strong preference for posts about specific topics or posts with specific styles and formats. We initialized the technology community with 500 random posts sampled from the subreddit r/technology. We then selected four groups of users based on unsupervised clustering of vote patterns, which we label as: (1)~Users who prefer human-centered content, e.g. technology that benefits people's health or can be applied to daily life, or labor rights in large technology companies; (2)~Users who prefer content related to security and regulation, e.g. cybersecurity and software security, government guidance and misinformation; (3)~Users who have broad interests, i.e., users who upvoted on a wide variety of topics; (4)~Users who prefer user-generated content (e.g. discussion posts), instead of posting a link to other technology media. In Table~\ref{tab:technology_curator_content}, we observe that content curated by each curator group again strongly aligns with the tastes of the curators.

\subsection{Online experiment}

Does curation produce shifts that are strong enough to be recognizable and distinguishable from current community approaches? %

To answer this question, we recruited $N=20$ participants from Amazon Mechanical Turk with a Masters qualification who had a 97\% approval rate and at least 5000 approved HITs, and paid them \$7.50 each for a thirty-minute study. Participants were presented with fifteen pairs of feeds. Each pair presented two versions of a feed for a Reddit subreddit, one of which was curated by a target group, and participants' goal was to identify which of the two versions was curated by the target group. The correct feed was curated\footnote{To enable more diversified curator selection for the experiment, we trained our curation model on a larger dataset. The larger dataset includes every vote on posts in the following set of popular subreddits on the topics of politics (r/politics, r/PoliticalDiscussion, r/Conservative, r/Liberal, r/Republican, r/democrats, r/VoteBlue), jokes (r/Jokes, r/Showerthoughts), science (r/science, r/ScienceFacts, r/technology, r/shittyaskscience), interest and ideologies (r/gaming, r/tattoos, r/MakeupAddiction, r/Music, r/punk, r/Fitness, r/travel, r/programming, r/hacking, r/Bitcoin, r/conspiracy, r/Futurology), movies and genres (r/movies, r/RomanceBooks, r/Marvel, r/marvelstudios, r/scifi, r/sciencefiction, r/Drama, r/anime, r/Documentaries, r/StarWars), lifestyle (r/teenagers), world, countries and regions (r/worldnews, r/britishproblems, r/europe, r/france, r/unitedkingdom, r/canada, r/australia, r/india, r/Philippines), religions (r/atheism, r/Christianity, r/Buddhism, r/islam), and gender, sexual orientation, and identity (r/Feminism, r/MensRights, r/LesbianActually, r/gay, r/trans).} by the stated group, and the incorrect feed was generated either by typical community upvoting (broadcast), or through curation by a different group. Example comparisons included: r/technology vs. r/technology curated by members who are also members of r/programming, r/Jokes vs. r/Jokes curated by members who are also members of r/teenagers, and r/worldnews curated by members who are also members of r/india vs. r/worldnews curated by members who are also members of r/france. Both pairs of feeds were generated by sampling the same five hundred posts from the subreddit, then ranking. Any traditional upvote (broadcast) feeds were ranked by the posts' actual Reddit score (upvotes - downvotes), and any curated feeds were ranked by the number of upvotes predicted by the curation model. The threshold for both prediction confidence and curator upvote rate were set at 0.5. Feeds were limited to the top fifteen ranked items. Participants were blind to which version of the feed was which. To avoid confusing participants with a description of the curation algorithm, we asked participants to select which of the feeds was created via upvotes from the target group. For more information on the fifteen pairs of feeds, please refer to the \autoref{study_details}.

\begin{figure}[tb]
  \centering
  
  \includegraphics[width=\textwidth]{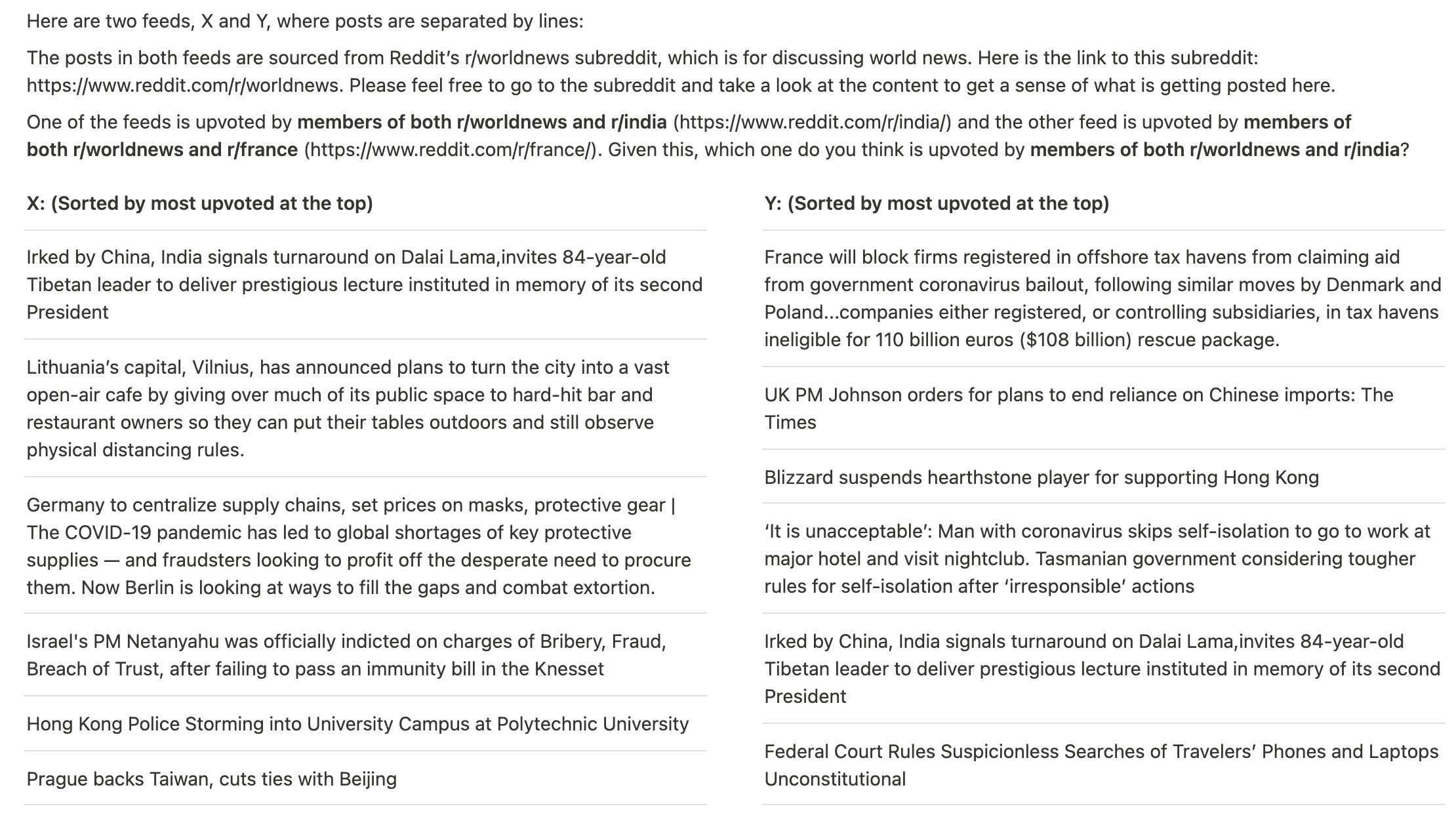}
  
  \caption{An example pair of feeds for comparison in the online experiment. To avoid confusion that might stem from a description of the curation algorithm, we asked participants to select the feed ``upvoted'' by a target group of community members.}
  \label{fig:teenager_subreddit_curators}
\end{figure}

For each pair, participants were told the subreddit and the curators, and were asked to guess which of the pair was generated by the curators as well as provide a written justification for their choice. To ensure the participants fully understood our task, we presented the participants with a training example, as well as the correct choice, at the start of the task.

We observed that 84.7\% of participants' selections were correct, far above a 50\% random guessing baseline. A one-proportion z-test ($N=20\times15=300$) confirms that participants recognized the curated feed at above random chance rate (z=$9.486$, $p<0.001$). These results suggest that curated feeds successfully execute recognizable shifts to the content of the community.

Participants were most successful when the curation group had distinct enough taste, and enough submissions in the subreddit representing that taste, to be recognizable. For example, every participant distinguished r/worldnews curated by r/india from r/worldnews curated by r/france, mentioning that the resulting feed explicitly mentioned ``India'' multiple times. In contrast, one of the more challenging comparisons was r/worldnews vs. r/worldnews curated by r/Liberal, likely because Reddit skews liberal and so the r/worldnews feed already embeds this point of view as a result. Future work in Cura could thus visualize for community administrators and moderators how much the curators' taste is similar to or different than the submission pool, general upvote pool, and other metrics, to understand what role curation is or is not playing.

\begin{table}[tb]
  \footnotesize
  \def\arraystretch{1.5}
  \begin{tabular}{p{2.4cm}p{3.2cm}p{2.6cm}p{4.3cm}}
  \toprule
    \textbf{Human-centered} &
    \textbf{Security and regulation} &
    \textbf{Broad interests} &
    \textbf{User-generated content} \\ \midrule
  Yes, Americans can opt-out of airport facial recognition — here’s how &
    Google tracked his bike ride past a burglarized home. That made him a suspect. &
    With more than 50 million US subscribers, Netflix has finally surpassed cable TV &
    We need to make it clear to the FCC that we want uncapped Internet access, for innovation in an increasingly data dependent world and user protection. \\ %
  A Worker in Amazon's New York Warehouse Has Died of the Coronavirus &
    Facebook Will Ban Protests That Defy Government 'Guidance' on Distancing &
    Trump calls for 6G cellular technology, because why the heck not &
    Hey guys, Eric from Netflix, letting you know we're joining reddit and others for ‘Internet Slowdown’ Day Sept. 10th to protect Net Neutrality. \\ %
  Toshiba says its device tests for 13 cancer types with 99\% accuracy from a single drop of blood &
  In 2020, Some Americans Will Vote On Their Phones. Is That The Future? - For decades, the cybersecurity community has had a consistent message: Mixing the Internet and voting is a horrendous idea. &
    Robocallers blasted Americans with 26.3 billion spam calls last year - Robocalls are up 46 percent from 2017 &
    If Google, Facebook, Twitter, and Yahoo really want to raise awareness on SOPA, they should follow Wikipedia's idea and shut down their sites and services for the day. TL;DR Goo, FB, Twit, and Yaho should do a wiki. \\
  A Device That 'Prints' New Skin Right Onto Burns Just Passed Another Animal Trial &
  Macs now twice as likely to get infected by adware than PCs, according to research &
    Alicia Keys using sealable pouches to lock up concert goers phones to have a 'phone free show' &
    Eight members of Congress that voted to kill broadband privacy are now leading the charge to kill Net Neutrality as well  \\ \bottomrule %
  \end{tabular}
  \caption{Distinctive posts from the technology community preferred by different groups of curators (either only preferred by one group of curators or ranks much higher in one curated feed than others by curator upvote rate).
   }
  \label{tab:technology_curator_content}
  \end{table}

\subsection{Curation reduces anti-social content} %
Moderation requires substantial effort from volunteers to maintain pro-social norms, yet still one in twenty posts on Reddit post moderation contains violations such as misogyny, personal attacks, or bigotry~\cite{park2022measuring}.
Here, we demonstrate that curation dramatically reduces the rate of anti-social content without requiring any additional effort from moderators.
By selecting curators who do not upvote norm-violating content, such content is not curated into the community's feed.

We focus on the challenging context of r/teenagers, following the previous method of curator selection, creating conditions for broadcast (raw Reddit score), democratic curation (modeling all users who have at least five upvotes in r/teenagers), as well as four conditions curating r/teenagers via its users who also actively upvote content in r/trans, r/gay, r/LesbianActually, or r/Feminism. We sample $1500$ random post submissions from r/teenagers and then measure the macro-norm violation rate of the subset of these posts curated in each condition with a curator vote threshold of $50\%$ by replicating the AI+crowd worker annotation pipeline for identifying Reddit macro-norm violations\footnote{A post is regarded as norm-violating if it violates any of the eight macro-norms identified by \cite{park2022measuring}, including using misogynistic or vulgar slurs, inflammatory political claims, bigotry, verbal attacks on Reddit or specific subreddits, posting pornographic links, personal attacks, abusing and criticizing moderators, or claiming the other person is too sensitive.} developed by \citeauthor{park2022measuring}~\cite{park2022measuring}. For the crowd annotation, we enlist United States-based Mechanical Turk workers with a Masters qualification who have 97\% approval rate and at least 5000 approved HITs. Three workers review whether a particular post violated each of the macro-norms from prior work, and we label each post as a violation if two of the three workers labeled it as violating at least one macro-norm.

\begin{figure}[tb]
  \centering
  \includegraphics[width=0.56\textwidth]{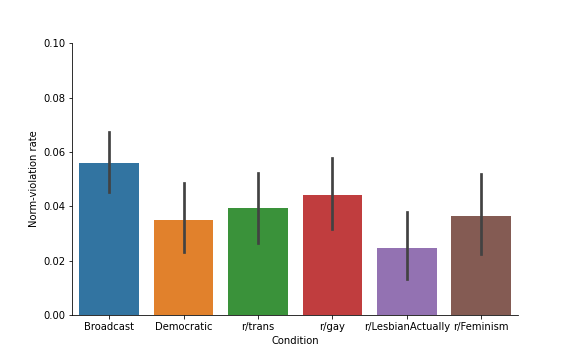}
\includegraphics[width=0.35\textwidth]{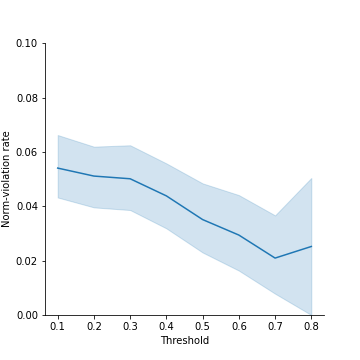}
  \caption{(A) Norm-violation rate of broadcast ($5.60\%\ [4.49, 6.89]$), democratic curation ($3.51\%\ [2.36, 5.0]$) and multiple small group curation. (B) Under democratic curation, as the threshold for curator upvote rate increases from $0.1$ to $0.8$, the norm-violation rate decreases from $5.41\%\ [4.31.6.68]$ to $2.52\%\ [0.00,5.43]$ (the small increase is due to the bias caused by few data points). The error bar represents the confidence interval.
  }
  \Description{Figure A displays a bar chart of norm-violation rate and its confidence interval for broadcast condition (5.60%
  Figure B displays a line chart where the x-axis is the threshold for curator upvote rate, and the y-axis is the norm-violation rate. As the threshold for curator upvote rate increases from 0.1 to 0.7, the norm-violation rate decreases from 5.41%
  }
  \label{fig:norm_violation_rate}
\end{figure}

We first replicate prior work by finding that the broadcast condition features one in twenty posts violating at least one macronorm: $5.60\%$, 95\% CI $[4.49, 6.89]$. The curated communities substantially reduce this rate, with no extra manual moderation effort. Democratic curation---the simplest curator selection---decreases the norm-violation rate to $3.51\%\ [2.36, 5.0]$ (\autoref{fig:norm_violation_rate}(A)). Better results can be further achieved by selecting appropriate community members as the curators: the community curated by users who actively upvote in r/LesbianActually achieves a norm-violation rate of $2.48\%\ [1.39,4.05]$, reducing the overall rate by over half. A Chi-Square test confirms that the difference between conditions is significant ($\chi^2(5)=13.4$, $p<.05$).

Setting different thresholds for the curator upvote rate results in different front-stage feeds as well (\autoref{tab:teenager_threshold}). High thresholds (e.g., $0.8=80\%$ of curators upvoting a post in order to be curated) encourage positive posts and sharing achievements, $0.6$ includes more humor and joke posts, and $0.2$ shows weak control over content, allowing diverse values and negative posts. 
We again measure the norm-violation rate under democratic curation, this time varying the curation threshold, and find that the norm-violation rate significantly decreases as the threshold increases, down to again roughly 2\% (\autoref{fig:norm_violation_rate}(B)), indicating that higher thresholds encourage tighter norm adherence and reduce harm.

\subsection{From broadcast to lurker-inclusive democracies} %
Compared to broadcast, which only amplifies the voices of users who vote, communities can build more democratic spaces by selecting a wider set of users as the curators, or create a more centralized community by selecting a small group with distinctive taste. %
To evaluate the effect of switching to a more democratic community, we conduct an experiment on r/teenagers. The \textit{broadcast} condition ranks by Reddit's raw score (upvotes-downvotes), which effectively only considers actively voting users. The \textit{small group curation} condition follows the same method as before by selecting different sets of curators as users who upvote content both in r/teenagers and r/Feminism, r/LesbianActually or r/gay.
Finally, the \textit{democratic curation} condition includes all participants in the community with $\geq 5$ votes in the community (i.e. those who are at minimally active and likely understand the community's norms) as curators.

\begin{table}[tb]
  \begin{tabular}{l|rrrrrr}
                   &  Broadcast & r/Feminism  &  r/LesbianActually & r/gay  & Democratic \\ \hline
Broadcast          &    1. \\ 
r/Feminism         &    0.503   &    1. \\ 
r/LesbianActually  &    0.585   &    0.686    &         1. \\ 
r/gay              &    0.611   &    0.760    &         0.772      & 1. \\ 
Democratic         &    0.664   &    0.812    &         0.841      & 0.919  & 1. \\ 
  \end{tabular}
  \caption{Different conditions result in content with very different ranking. This table reports Spearman's rank correlation coefficients of the feed ranking across broadcast, small group curation and democratic curation.}
  \label{tab:curator_spearman}
\end{table}

We then measure the Spearman's rank correlation coefficient between the resulting ranked feeds (\autoref{tab:curator_spearman}).
Broadcast and democratic curation feature a rank correlation of $0.664$, suggesting that post rankings might change substantially between Reddit today and a democratically curated system. Democratic curation combined the preferences of multiple small groups of curators: the rank correlation between democratic curation and curation by curators from r/Feminism, r/LesbianActually or r/gay are all above $0.81$.

\section{Discussion}
In this section, we reflect on our contributions and limitations. We discuss how communities might use the Cura system in practice, and discuss potential risks and ethical considerations of our approach.

\subsection{Curation, broadcast, and the rules of social media platforms}
Socio-technical systems shape the behavior of the people in them, and the systems are shaped by the people in them~\cite{ackerman2000intellectual}. These systems embed rules that either directly control or indirectly influence who can speak, in what ways, and when.

Online community software today has largely aligned around the design metaphor of \textit{broadcast}. In a broadcast metaphor, platforms focus on empowering users to share content instantaneously: sending posts or replies to all others in a community, to a newsfeed, or to all followers, before inspection. Research has examined how to encourage participation~\cite{burke2009feed}, shape newcomer behavior~\cite{matias2019preventing}, and manage anti-social behavior~\cite{kiesler2012regulating} in systems with such a broadcast metaphor.

Curation sits in a space of alternative design metaphors for social media. Could social media provide more social translucence~\cite{gilbert2012designing}, stronger notions of consent~\cite{im2021yes}, or other rules that are not tethered to or trying to fit within the broadcast metaphor~\cite{hollan1992beyond}? We explore curation here because it inverts the logic of instant broadcast, providing trusted parties with more levers to shape the social space, and trades off content quantity for control. Future research can continue to explore this and other metaphors to offer a more varied palette of social designs for communities to select from.

\subsection{Application in practice and implications for design}
Cura is currently an algorithmic and visualization layer on top of Reddit data. Cura would require a community to change its back-end voting, or install a bot to do this on its behalf, to fully instantiate curation on existing platforms, not a small change. Still, in terms of the user experience of reading and upvoting posts, Cura is not far from existing social media platforms such as Reddit. In principle, Reddit or Facebook Groups could enable groups to nominate curators and then offer a separate ranking view for curation. Moreover, curation can also be combined with a personalized feed: curation can serve as a common filter prior to personalized ranking. While curation determines what is the inventory of the posts publicly visible to the community, personalized feed ranking can determine the subset of the inventory presented to the end user.

Automated moderation algorithms offer another possible horizon for this approach. Current moderation algorithms focus on simple term lists or regular expressions, yet often falsely label appropriate content as inappropriate. It is possible that variants of this approach might offer an alternative that can generalize based on learnable model parameters, or for example the ability to predict whether curators will block a piece of content.

Social media platforms that are composed of multiple communities with different topics and tastes, e.g., Reddit, are the best fit for curation as described. 
For platforms without community structure, e.g., Twitter and Facebook, the implementation would need to be adjusted, perhaps by considering the whole platform as a global community, where curation can help to maintain the global norms in the platform. Alternatively, in the case of Twitter communities and Facebook Groups, where the platform has implemented communities for users to share information, ideas, and experiences related to a specific topic, Cura can be immediately applied. By selecting experienced members of the groups as curators, Cura can provide a more focused vision by selectively displaying curated discussion threads. 

In practice, it may be more straightforward to launch a new community platform for curation. Doing so would help portray a simpler mental model to users, where it is clear that all communities are curated, and the voting user interface can be better customized to communicate curation thresholds.

More broadly, our work is a reaction to social computing policy and algorithmic solutions that assert that communities and platforms can only perform \textit{downstream} reactions and clean-ups after anti-social behavior has occurred. As the social computing research community has demonstrated, however, a more effective approach may be to design \textit{upstream} changes that shift behavior so that the anti-social behavior never arises in the first place. Our work serves as a viable alternative to these reactive solutions, and other design alternatives are also available. For instance, a platform may consist of hierarchically connected communities that enables high-quality and cross-cutting content to traverse the network.

There may be concerns regarding the potential negative impact of Cura on user engagement, which might further lead to degradation in model performance. However, we posit that users are still motivated to vote as before, given that their votes can improve the performance of the curation model and have an impact on the outcome. While there may be instances where a user's vote holds minimal weight due to behaviors such as trolling, the potential secondary effect that they vote less still bring about positive benefits for the community.

\subsection{Scaling from small to large communities}

Cura is not a good fit for every community. As a general gideline, for a small, tightknit community that contains less than 50 people who know and highly trust each other, little curation may be necessary---the community may be a better fit for broadcast. In this situation, the curation model may also be less accurate, given less voting data for model training. If the community grows, however, this can provide valuable training data for the curation model and allow the community to engage with curation immediately. For medium-sized communities consisting of 50 to 500 users, we recommend utilizing Cura. In the case of large communities with over 500 users that desire a curated flavor, we strongly recommend adopting Cura to support the community.

\subsection{Governance}\label{sec:governance}
Curation is compatible with new modes of online governance~\cite{zhang2020policykit}. For instance, platforms could set up macro-level norms that are shared by all communities, place global training restrictions on what the curation model could learn to accept, and what content the model should predict to downvote on regardless of the user it is simulating. Or, a platform-wise curation could be applied prior to commmunity-wise curation, and positions as curators could be democratically elected by each community. %

Curator selection is essential to make the most of curation: it is decisive for the taste and values of a community. Depending on the curators selected, groups may use Cura to support certain viewpoints or suppress other viewpoints, including those from minorities and historically disadvantaged groups.
Existing discussions on who and how to curate content---including social media posts, news feeds and art---have considered professional expertise~\cite{rosenbaum_can_nodate}, diversity~\cite{Gaupp+2021+290+321}, perspective diversity~\cite{Cui2017HowDO} and other values. Based on our experience in this project, we reflect on four important dimensions that should be taken into consideration during curator selection: 

\begin{enumerate}
    \item \textit{Democracy.} Is the community selecting all its members as the curators, or only a small set of members as the curators, e.g., only moderators or only the active users? Additionally, democratic values can manifest within the selection of curators: do the community members get to vote or otherwise influence curator choice?~\cite{frey2019place,zhang2020policykit}
    \item \textit{Expertise.} Are the curators experts in the topic/field of the community? Expert curators can help support high quality community content, but this is in tension with democratic communities.
    \item \textit{Opinion diversity.} Do the curators hold differing viewpoints, or do they have very similar taste? Curators with high opinion diversity can help prevent echo chambers. A community should take into consideration, though, if the curators' preferences have muddled or even strongly conflicting, performing curation may lead to a community without clear norms or purpose. 
    \item \textit{Identity diversity.} Are the curators a balanced selection of users with different identities and from different stakeholders groups~\cite{gordon_jury_2022}? 
\end{enumerate}

We allow administrators to select curators for simplicity of explanation, but in many communities, they might want curators to be selected through a democratic election. Communities may also explore other possible procedures for selecting curators and different forms of governance~\cite{zhang2020policykit}.

The current Cura system regards every curator equally---the final curation decision is based on the aggregation of one-person one-vote. However, consider editors in media: there are usually multiple editors with one chief editor whose opinion is decisive. Or consider the fan community of a celebrity, with the celebrity themselves also being a member: the voice of the celebrity themselves usually account more than other fan members.
In these situations, it may be appropriate to weigh voices unequally.
Future research could develop an improved interface that enables administrators to transparently manage their curators and assign different weights to them. The interface could serve as a ``staging'' area, where administrators can experiment with different weights and balances of curators. They can compare side-by-side feed simulations of different settings to understand the impact of those changes on the community before implementing them live.

\subsection{Ethical considerations}

In curation, community norms and values are explicitly defined by naming curators. Today's communities still follow the taste of certain users, but that decision is instead implicit, based on the platform design and relative activity level of different community members. Our normative position is that it is better to force this value-laden decision to be made explicitly, where it can be seen and deliberated, rather than implicitly allowing anti-social behavior to persist under the cover of decentralized community upvoting as in today's platforms. Making this choice explicit does not guarantee fairness or equity; but it gives the community members a better chance to make informed decisions about the communities they invest in.%

However, increasing visibility on the curators' votes can cause harm; marginalized community members, possibly selected for their unique perspective, may end up as targets of harassment from those who dislike their curatorial decisions, and displaying their decisions can add fuel to the fire. In practice, we suggest notifying curator candidates about the potential consequences of being selected curators and asking them for affirmative consent before defining them as curators, letting curators' votes be audited, but not necessarily visible publicly so curators have plausible deniability and are not discouraged from expressing their opinions. We also suggest having several curators for contentious groups, so that the curation result is their combined effort and no one becomes a lone target.

Curation centralizes power in the hands of the curators, which is not a value-neutral decision. Centralized power has historically been used to oppress and suppress undesired viewpoints. An approach to mitigating this risk is enabling centralized curation, but requiring that curator selection be subject to democratic election: a technique advocated for in the feminist essay ``The Tyranny of Structurelessness''~\cite{freeman1972tyranny}.

How do we stop malicious actors from creating communities that curate anti-social or evil content? %
Platforms can mitigate the risk of this outcome, for example, by placing alarms or restrictions on the kinds of content that a curation model is allowed to learn to upvote. This could be achieved through a hidden test set of objectionable material: if a curator is chosen such that the model would upvote the objectionable material, the platform may refuse to allow that curator, or refuse to allow the model to upvote such material.

Another concern is strategic manipulation of the algorithm, e.g., upvoting toxic content to affect the model's prediction and push specific content to the frontstage. Such behaviors by a small number of people will not change the model's prediction (\autoref{fig:peer_vote_support_adv}(B)). The system also has some measure of incentive compatibility: if the ill-intentioned users are continuously trying to trick the algorithm, the system will learn that their votes are uncorrelated with the curators' opinions, thus those users will gradually lose their ability to have any impact on the prediction. A more serious threat could involve an individual disguising themselves as a respected community leader and subsequently manipulating voting patterns once selected as a curator by the administrator. Future research could develop automated systems that can detect possible attackers by analyzing inconsistencies in their voting behavior and identifying significant discrepancies from other users' voting histories.

\subsubsection{Audits and oversight}
Curation models can be audited: an automated audit could regularly test and identify which communities' models are letting hateful content through, and quarantine or deplatform those communities~\cite{chandrasekharan2017you,chandrasekharan2022quarantined}. 
As mentioned above, platform implementation could also place global training restrictions on what such models could learn to accept, making it difficult for communities to post any hateful content. So, if combined with platform governance, curation models may offer some opportunities here as well.

\subsubsection{Filter bubbles and echo chambers}
One might worry that curation will harm the content diversity of a community, only presenting content that are from the same perspective, leading to filter bubbles and echo chambers. The community might reinforce extreme and unhealthy behavior, e.g., a beauty community encourages an unhealthy pursuit of beauty through excessive dieting.
To mitigate these risks, a few interventions are possible:
\begin{enumerate}
\item Systems might help users visualize the extent to which their communities represent a broad swath of points of view~\cite{munson2013encouraging}.
\item Communities' administrators might diversify curator selection. A community can have ``balancing forces'' in curator seats, of the sort that a representational democracy offers, if thought through carefully. %
\item Communities might be connected hierarchically so that no community is an island unto itself: each community's content can percolate up from the metaphorical neighborhood to city to county to country, and vise versa, ensuring that cross-cutting content traverses the network. 
\end{enumerate}

However, the same features that invoke fears around echo-chambers also facilitate safe and purposeful spaces; the interventions mentioned above are not solutions for every community. In contrast to the example of a beauty community above, imagine the case of a community for eating disorder recovery, where not only would opposing viewpoints be directly harmful, but content from other communities trickling in (for example, high fashion or modelling) could be as well. These contrasting examples show that there is no single correct decision for all contexts; rather, there are many thoughtful decisions to be made when applying the curation process to different communities.

\subsection{Limitations}
While we perform technical evaluations and demonstrations of curation, this work does not yet report a field deployment. As a result, we cannot report second-order behavioral effects of curation on a community, nor how its members react to curation. Future work can investigate the emergent effects of curation on community behavior and norms: does it create more norm-adherent behavior? What proportion of content makes it over the curation threshold? Are subjective experiences of curation enjoyable, or frustrated?

Since the present Reddit dataset was sourced from Reddit users who voluntarily made their voting data public, it is possible that our dataset includes a higher proportion of frequent Reddit users, who exhibit above-average rates of voting. Nevertheless, if implemented on a real platform where complete voting data is accessible, the Cura system would have fuller access to all voting behavior that users are willing to share with the system.

The Reddit dataset we use is collected from real users, but not from communities that were originally intended to be curated. So, our results may differ if communities and votes were initially authored with the express purpose of curation. Likewise, Reddit users who voluntarily make their vote data public may differ categorically from other Reddit users, potentially limiting the generalizability of our findings. Since the original dataset is also extremely large and easily exhausts many computational resources, we currently can only make claims about a subset of sampled subreddits. %
While we restrict our curators to users who have voted at least $5$ times, the curation model needs to be finetuned or retrained before being applied to very small communities. Another limitation of our dataset is that it lacks the image and video data in Reddit posts, presenting great challenges for our model to predict curators' votes on posts where images or videos are the main content.

The AI models underlying curation are not perfect. Perfect prediction is impossible, since opinions are never fully predictable and can be subject to social influence. So, it is important that curators can vote manually to override the model's prediction for them, and that the models retrain with curators' votes to improve their performance. However, perfect prediction is also not necessary for curation to succeed, given the inherent noise around social ranking~\cite{gilbert2013widespread}.
\section{Conclusion}
We present Cura, a system that enables curation in social media with algorithms and interface.
Curation is a metaphor in which the curators of a community decide which content is shared with the larger group, placing trust in that curator's taste to maintain the norms and content quality of the group. Although commonly applied in news media, requiring curators to review every piece of content is too effort-intensive to enable it being instantiated in social media. 
We overcome this barrier by leveraging a transformer-based deep learning model to predict whether each curator would upvote each post, based on community feedback.
We evaluate our approach and demonstrate that our model can accurately estimate curators' opinions and updates quickly as votes from members arrive. We also demonstrate that curation can inflect the same inventory of posts into many different community types, depending on the curators' taste. Curation enables a wide variety of community types ranging from a small group of editors (e.g., newspaper editorial boards), to stakeholder roundtables (e.g., including minoritized groups as curators), to democracies (e.g., giving every user equal voice, instead of just the small number of active users or moderators).

\begin{acks}
This work was supported by the Office of Naval Research. Mitchell Gordon was supported by the Apple Scholars in AI/ML program.

We are grateful to the Chinese Undergraduate Visiting Research Program for providing Wanrong He with the opportunity to undertake this research internship at Stanford University.

We would like to thank the participants of our online experiments, whose feedback greatly contributed to the findings of this research. Their engagement was crucial in shaping the outcomes and enhancing the overall quality of this work.

Additionally, we acknowledge and thank Kris Jeong, Nicole Garcia, and Pauline Arnoud for their collaboration and teamwork throughout this project. They brought fresh ideas and perspectives to the table, broadening the scope and depth of this research.

Finally, we express gratitude to all the individuals, mentors, and colleagues who provided guidance and assistance throughout this research. Their advice, encouragement, and expertise were indispensable in shaping the direction and outcomes of this work.
\end{acks}

\bibliographystyle{ACM-Reference-Format}
\bibliography{acmart}

\appendix 

\section{Online Experiment Details} \label{study_details}
We list how the fifteen pairs of feeds in our online experiment are generated in \autoref{tab:feed_detail}, including the Reddit subreddit being used, and the curator groups that curate the target feed and the incorrect distractor feed, as well as if the distractor feed is generated through typical community upvoting (broadcast) or curation.

\begin{table}[]
\begin{tabular}{p{6.5cm}p{6.5cm}}
\toprule
\textbf{Target Feed}                                                          & \textbf{Distractor Feed} \\ \midrule
r/technology curated by members who are also members of r/programming & r/technology   \\ \midrule 
r/technology curated by members who are also members of r/teenagers & r/technology curated by members who are also members of r/Conservative  \\ \midrule
r/PoliticalDiscussion curated by members who are also members of r/Conservative & r/PoliticalDiscussion  \\ \midrule
Super-politics community (sampling 500 posts from each of r/politics, r/Conservative, r/Liberal, r/Republican, r/democrats, and r/PoliticalDiscussion) curated by members who are also members of r/democrats & Super-politics community curated by members who are also members of r/Republican  \\ \midrule
r/Jokes curated by members who are also members of r/LesbianActually & r/Jokes  \\ \midrule
r/Jokes curated by members who are also members of r/teenagers & r/Jokes   \\ \midrule
r/Jokes curated by members who are also members of r/Conservative & r/Jokes  \\ \midrule
r/teenagers curated by members who are also members of r/gaming & r/teenagers   \\ \midrule
r/teenagers curated by members who are also members of r/travel & r/teenagers curated by members who are also members of r/punk   \\ \midrule
r/worldnews curated by members who are also members of r/Liberal & r/worldnews   \\ \midrule
r/worldnews curated by members who are also members of r/india & r/worldnews curated by members who are also members of r/france   \\ \midrule
r/gaming curated by members who are also members of r/teenagers & r/gaming   \\ \midrule
r/gaming curated by members who are also members of r/LesbianActually & r/gaming curated by members who are also members of r/scifi   \\ \midrule
r/music curated by members who are also members of r/Christianity & r/music curated by members who are also members of r/scifi   \\ \midrule
Super-science community (sampling 500 posts from each of r/science, r/ScienceFacts, r/technology, and r/shittyaskscience) curated by members who are also members of r/programming & Super-science community curated by members who are also members of r/Jokes   \\ \bottomrule
\end{tabular}
\caption{Generation details of the fifteen pairs of feeds.}
\label{tab:feed_detail}
\end{table}

Also note that the training example is:  r/PoliticalDiscussion curated by members who are also members of r/democrats (correct) vs. r/PoliticalDiscussion curated by members who are also members of r/Republican (incorrect).

\end{document}